\documentclass[useAMS]{mn2e}
\usepackage{times,psfig,epsfig}

\newcommand{\hm}{\,h^{-1}{\rm Mpc}}
\newcommand{\chandra}{{\it Chandra} }

\newcommand{\xmm}{{\it XMM-Newton} }

\title[Evolution of the X-ray properties of simulated galaxy clusters]
{Evolution at $z \ge 0.5$ of the X-ray properties of simulated galaxy 
clusters: comparison with the observational constraints}

\author[S. Ettori et al.]
{S. Ettori$^1$, S. Borgani$^{2,3}$, L. Moscardini$^4$, G. Murante$^5$,
  P. Tozzi$^6$, \\~\\
\LARGE{\rm A. Diaferio$^7$, K. Dolag$^8$, V. Springel$^9$, G. Tormen$^8$,
 L. Tornatore$^2$} \\~\\
\footnotesize 
$^1$ ESO, Karl-Schwarzschild-Str. 2, D-85748 Garching, Germany 
(settori@eso.org) \\
$^2$ Dipartimento di Astronomia dell'Universit\`a di Trieste, via
  Tiepolo 11, I-34131 Trieste, Italy (borgani,tornatore@ts.astro.it)\\
$^3$ INFN -- National Institute for Nuclear Physics, Trieste,
  Italy\\ 
$^4$ Dipartimento di Astronomia, Universit\`a di Bologna, via Ranzani
  1, I-40127 Bologna, Italy (lauro.moscardini@unibo.it)\\
$^5$ INAF, Osservatorio Astronomico di Torino, Strada Osservatorio 20,
  I-10025 Pino Torinese, Italy (murante@to.astro.it)\\
$^6$ INAF, Osservatorio Astronomico di Trieste, via Tiepolo 11,
  I-34131 Trieste, Italy (tozzi@ts.astro.it)\\
$^7$ Dipartimento di Fisica Generale ``Amedeo Avogadro'', Universit\'a
  degli Studi di Torino, via Giuria 1, Torino, Italy (diaferio@ph.unito.it) \\
$^8$ Dipartimento di Astronomia, Universit\`a di Padova, vicolo
  dell'Osservatorio 2, I-35122 Padova, Italy (kdolag,tormen@pd.astro.it)\\
$^9$ Max-Planck-Institut f\"ur Astrophysik, Karl-Schwarzschild Strasse
  1, Garching bei M\"unchen, Germany (volker@mpa-garching.mpg.de)\\
}

\date{submitted on 23 Feb 04; accepted on 1 Jul 04}

\begin{document}
\maketitle 

\begin{abstract}
We analyze the X-ray properties of a sample of local and high 
redshift galaxy clusters extracted from a large
cosmological hydrodynamical simulation. This simulation has been 
realized using the Tree+SPH code {\tt GADGET-2} for a 
$\Lambda$CDM model. It includes radiative cooling, star formation and 
supernova feedback and allows to resolve radially the thermodynamic 
structure of clusters up to redshift $1$ in a way that is not yet 
completely accessible to observations. 
We consider only objects with $T_{\rm ew} >2$ keV to avoid the large
scatter in the physical properties present at the scale of groups
and compare their properties to recent observational constraints.
% We have selected 97 galaxy clusters at redshift 0, 83 at $z=0.5$,
% 81 at $z=0.7$ and 72 systems at $z=1$. 
In our analysis, we adopt an approach that mimics observations,
associating with each measurement an error comparable with recent
observations and providing best-fit results via robust techniques.
Within the clusters, baryons are distributed among (i) a cold neutral
phase, with a relative contribution that increases from less than 1 to
3 per cent at higher redshift, (ii) stars which contribute with about
20 per cent and (iii) the X-ray emitting plasma that contributes by 80
(76) per cent at $z=0 \; (1)$ to the total baryonic budget.  A
depletion of the cosmic baryon fraction of $\sim$7 (at $z=0$) and 5
(at $z=1$) per cent is measured at the virial radius, $R_{\rm vir}$,
in good agreement with adiabatic hydrodynamical simulations.
% The X-ray emitting gas mass--fraction grows as $r^{\eta}$ with
% $\eta \simeq 0.25$ for $r < 0.3 \times R_{\rm vir}$,
% independent of the redshift.
% In the ten most hottest simulated clusters, 
% the measured gas mass fraction at $R_{2500}$, $R_{500}$ and $R_{\rm vir}$
% is about 60, 70 and 75 per cent of $\Omega_{\rm b}/\Omega_{\rm m}$
% both locally and at $z=1$, with slightly lower values at higher redshift
% although within the measured scatter.
We confirm that, also at redshift $>0.5$, power-law relations hold
between gas temperature, $T$, bolometric luminosity, $L$, central
entropy, $S$, gas mass, $M_{\rm gas}$, and total gravitating mass,
$M_{\rm tot}$ and that these relations are steeper than predicted by
simple gravitational collapse.  A significant, negative evolution in
the $L-T$ and $L-M_{\rm tot}$ relations and positive evolution in the
$S-T$ relation are detected at $0.5 < z < 1$ in this set of simulated
galaxy clusters.  This is partially consistent with recent analyses of
the observed properties of $z \ga 0.5$ X-ray galaxy clusters.  By
fixing the slope to the values predicted by simple gravitational
collapse, we measure at high redshift normalizations lower by 10--40
per cent in the $L-T$, $M_{\rm tot}-T$, $M_{\rm gas}-T$, $f_{\rm
gas}-T$ and $L-M_{\rm tot}$ relations than the observed
estimates. This suggests that either the amount of hot X-ray emitting plasma
measured in the central regions of simulated systems is smaller than
the observed one or a systematic higher value than actually measured
of gas temperatures and total masses is recovered in the present 
simulated dataset.
\end{abstract} 
 
\begin{keywords}  
cosmology: miscellaneous -- methods: numerical -- galaxies: cluster: general 
-- X-ray: galaxies. 
\end{keywords}

\section{Introduction}

Only in the recent years, and after deep exposures obtained through
the new generation of X-ray telescopes like \chandra and \xmm, it has
been possible to start investigating systematically the physical
properties of galaxy clusters, at redshift higher than 0.4
(e.g. Holden et al. 2002, Vikhlinin et al. 2002, Ettori et al. 2004).

Galaxy clusters are believed to form under the action of the dark
matter gravity in the hierarchical scenario of cosmic structure
formation. They assemble cosmic baryons from the field and heat them
up through adiabatic compression and shocks that take place during the
dark matter halo collapse and accretion.  Simple self-similar
relations between the physical properties in clusters are then
predicted (e.g. Kaiser 1986, 1991, Evrard \& Henry 1991) since the
gravity does not have any preferred scale and the hydrostatic
equilibrium between X-ray (mostly originated through bremsstrahlung
processes) emitting gas and the cluster potential is a reasonable
assumption. Therefore, how the baryons distribute within the cluster
potential as function of cosmological time and whether or not scaling
relations between integrated physical properties still hold and,
possibly, evolve at high redshift, these are fundamental diagnostics
to validate the self-similar scenario and to study the formation and
evolution of X-ray galaxy clusters. A main role in assessing these
issues is played by cosmological hydrodynamical simulations that can
drive the analysis of observationally unresolved high redshift
systems.

To this purpose, we have analysed the predictions obtained at high
redshift from the large hydrodynamical cosmological simulation
presented and discussed in Borgani et al. (2004, hereafter Paper I),
that has been shown to reproduce at $z=0$ several observed properties
of X--ray clusters, such as the local temperature function, the
luminosity--temperature relation for systems hotter than 2 keV, and
the mass--temperature relation.
 
As a natural follow-up of that work, we study here the evolution of the
X-ray properties of the simulated galaxy clusters from $z=0$ out to
$z=1$.  This paper is organized as follows: in the next section, we
describe our dataset of simulated clusters and their physical properties;
in Section~3 we discuss how the baryons, and particularly
the X-ray emitting plasma, are distributed in the local and $z \ge
0.5$ systems. In Section~4, we introduce the X-ray scaling relations
and discuss their evolution at redshift $0.5-1$ with a direct
comparison with the observational results. Finally, we summarize
and discuss our results in Section~5.  

\begin{figure}
  \epsfig{figure=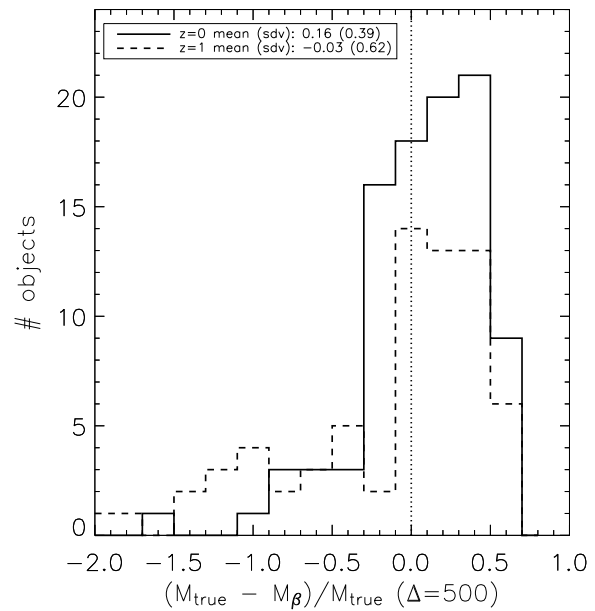,width=0.5\textwidth}
\caption{Comparison between the true cluster masses and 
estimates measured by using an isothermal $\beta-$model.
Mean deviations of about 15 per cent (at $z=0$) or less (at $z=1$) 
are obtained. 
} \label{fig:mass}
\end{figure}

\begin{figure*}
\hbox{
  \epsfig{figure=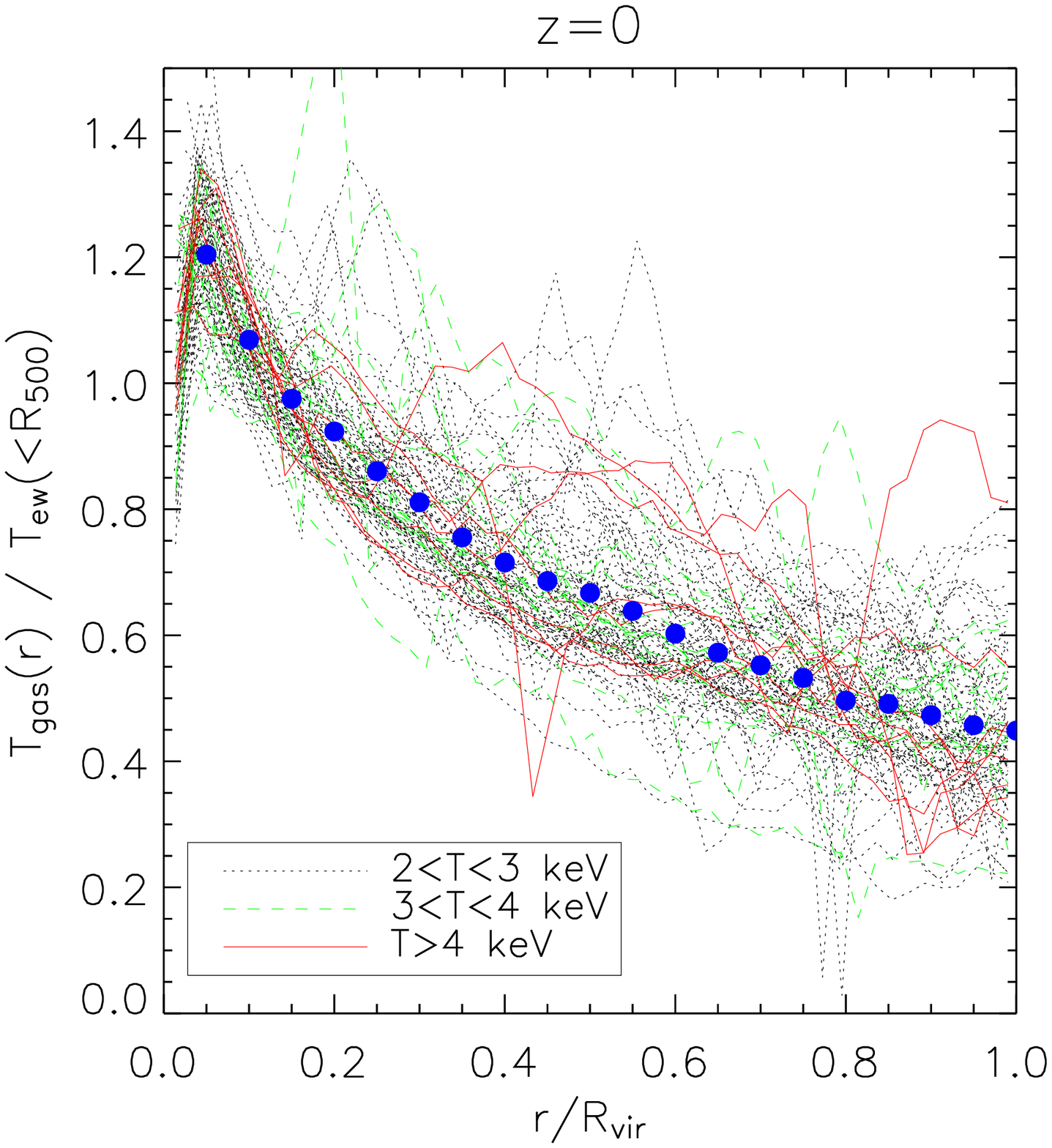,width=0.5\textwidth}
  \epsfig{figure=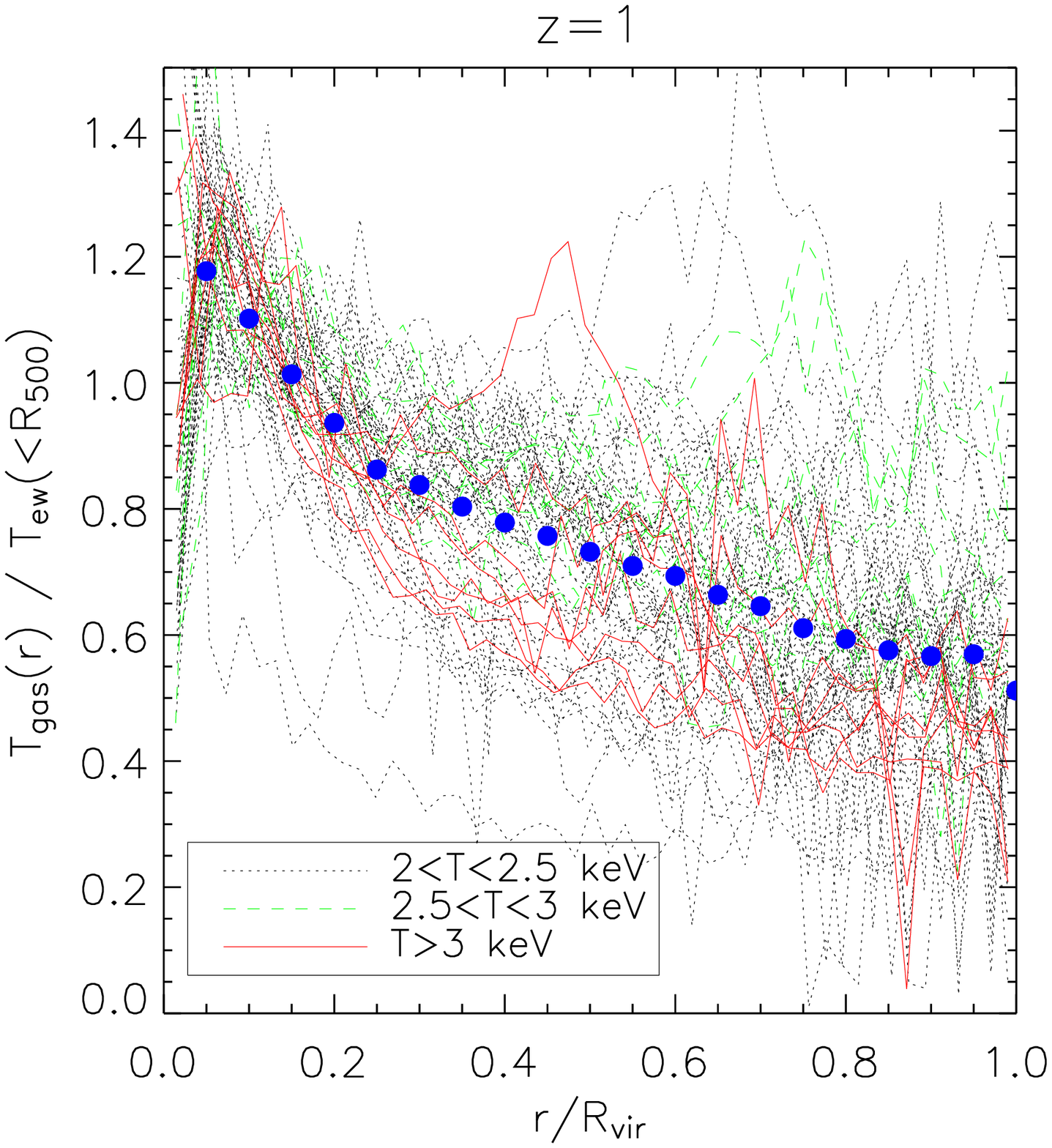,width=0.5\textwidth}
} \hbox{
  \epsfig{figure=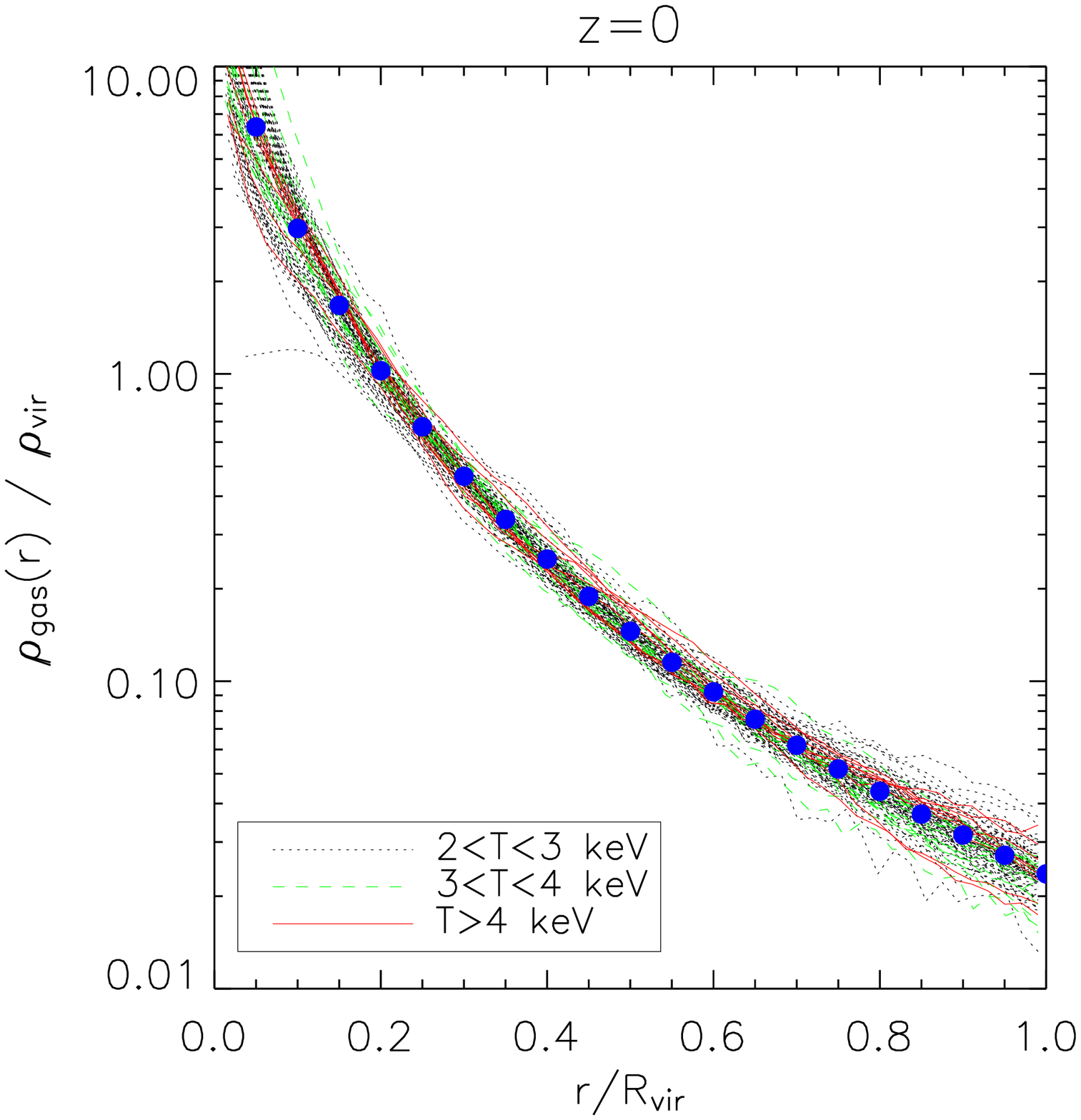,width=0.5\textwidth}
  \epsfig{figure=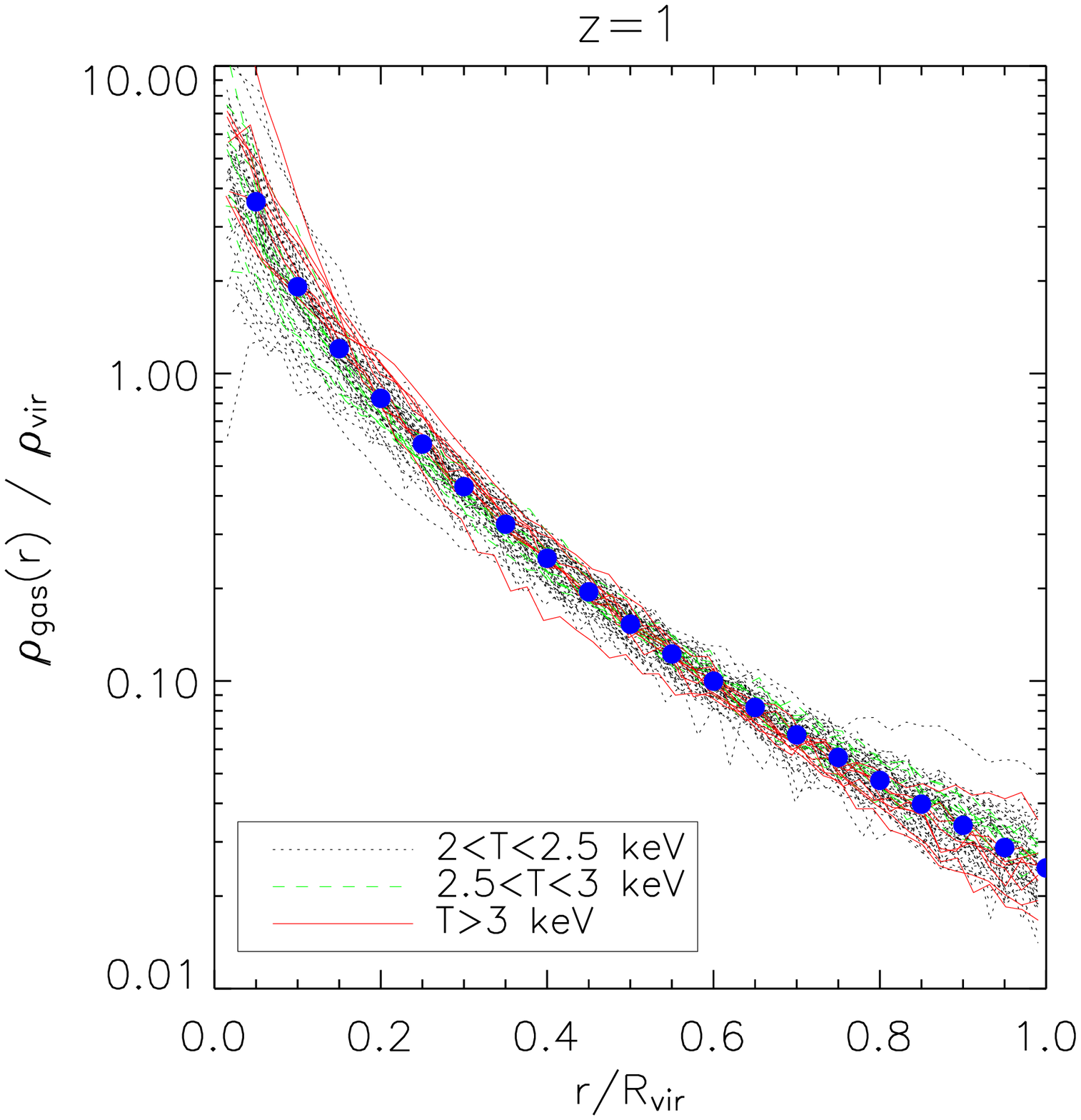,width=0.5\textwidth}
}\caption{All ({\it lines}) and average ({\it dots}) differential 3-D 
gas temperature (upper panels) and density (lower panels) profile 
at redshift $0$ (panels on the left) and $1$ (panels on the right).
The gas density is normalized to the virial value, $\rho_{\rm vir}$, 
that is defined as the mean density within 
$R_{\rm vir}$ and is equal to $\rho_{\rm c,z} \times \Delta_z$.
} \label{fig:tn_r}
\end{figure*}

\section{Properties of the simulated clusters}

The simulated clusters are extracted from the large--scale
hydrodynamical simulations described in Paper I. We refer to that
paper for a detailed description of the simulation, while we provide
here only a short summary. The simulated cosmological model is a
standard flat $\Lambda$CDM universe, with $\Omega_{\rm m} =
1-\Omega_{\Lambda} = 0.3$, $\sigma_8 = 0.8$, $\Omega_{\rm b} h^2 =
0.019$ and $H_0 = 100 h$ km s$^{-1}$ Mpc$^{-1}$ with $h=0.7$.  The
simulation follows the evolution of $480^3$ dark matters and an
initially equal number of gas particles within a box of $192\hm$ on a
side, so that $m_{\rm DM}=6.6\times 10^9M_\odot$ and $m_{\rm
gas}=9.9\times 10^8M_\odot$ for the mass of the DM and gas particles,
respectively. The Plummer--equivalent gravitational softening of the
simulation was set to $\epsilon_{\rm Pl}=7.5\,h^{-1}$ kpc comoving
from $z=2$ to $z=0$, while it was taken to be fixed in physical units
at higher redshift.

The run has been realized using {\small GADGET-2}\footnote{\tt
http://www.mpa-garching.mpg.de/gadget}, a massively parallel tree
N--body/SPH code (Springel, Yoshida \& White 2001) with fully adaptive
time--step integration, which includes the explicit energy and entropy
conserving SPH formulation by Springel \& Hernquist (2002).  The
simulation includes radiative cooling of a plasma of primordial
composition, the effect of a photoionizing, time--dependent, uniform
UV background (e.g., Haardt \& Madau 1999).  Star formation is treated
using the self--regulated hybrid multiphase model for the interstellar
medium introduced by Springel \& Hernquist (2003). The code also
includes a phenomenological description of galactic winds powered by
type--II supernovae, such that star-forming gas particles contribute
to the wind with a mass outflow rate two times larger than their star
formation rate, with a wind velocity of about $360\,{\rm km\,
s^{-1}}$.

Clusters are identified in the simulation box by first applying a
friends-of-friends halo finder to the distribution of DM particles,
with a linking length equal to 0.15 times their mean separation. For
each group of linked particles, we identify the particle having the
minimum value of the gravitational potential. This particle is then
used as the center of the cluster to run a spherical overdensity
algorithm, which determines the radius around the target particle that
encompasses a given overdensity.  In the following analysis, we will
consider a typical overdensity with respect to the critical density
estimated at redshift $z$, $\rho_{\rm c, z} = 3 H_z^2/ (8 \pi G)$, of
$\Delta_z = 500 \times \Delta_{c,z}/\Delta_{c,EdS}$, where
$\Delta_{c,z}$ is the ratio between virial and critical overdensity at
redshift $z$ for our cosmological model, while $\Delta_{c,EdS}$ is the
same quantity in an Einstein--de-Sitter cosmology. This overdensity is
chosen for convenience, since the regions within these overdensities
are the most well-studied ones from an observational point of view
(e.g. Ettori et al. 2004).  Hereafter, we indicate by ``$\Delta=500$''
the cluster regions enclosed within a sphere with radius $R_{500}$ and
total gravitating mass 
\begin{equation}
M_{500} = \frac{4}{3} \pi \rho_{\rm c, z} \Delta_z R_{500}^3.
\label{eq:mass}
\end{equation}  
We measure, on average, $R_{500}$ to be 0.63 times $R_{\rm vir}$ and 
constant in redshift (we define $R_{\rm vir}$ as the
radius encompassing an average density equal to the virial density for
our cosmology and at a given redshift; e.g., Eke et al. 1996, 
Bryan \& Norman 1998). This confirms that a fixed fraction of the 
virial radius is mapped by using the approximation to the solution 
for the collapse of a spherical top-hat perturbation. 
It is worth noticing that if one fixes the overdensity independently of 
the cosmological parameters and redshift, larger portion of the virial 
regions are mapped at higher redshift, since dark matter halos are more 
concentrated (i.e. have lower radial radius). E.g., assuming
$\Delta_z=500$, one obtains $R_{500}/R_{\rm vir} =$ 0.49, 0.56 and 0.58
at redshift 0, 0.5 and 1, respectively.

As we show in Paper I and discuss further below in this section, the
simulated gas temperature and density profiles present clear
deviations from the mean behaviour of the observed ones. Therefore, we
decide to adopt the mass as defined in equation~\ref{eq:mass} and
inferred from the simulated dark matter halos, instead of the
corresponding value measured, e.g., through the hydrostatic
equilibrium.  We have measured, however, the reconstructed mass
estimates at $R_{500}$ by using an isothermal $\beta-$model (Cavaliere
\& Fusco-Femiano 1976) with a single emission-weighted temperature and
inferring $R_{500}$ from the corresponding mass profile (this
procedure follows very closely what is generally done with
observational data, e.g. Ettori et al. 2004).  It is worth noticing
that the mass profile is then a direct function of the best-fit
modelization of the gas density profile that, as we discuss at the end
of this section, shows evident mismatches with what is typically
observed.  We obtain that the $\beta-$model estimates underestimate
the true cluster mass by about 15 per cent on average at $z=0$ and are
in good agreement at $z=1$ even though with a larger dispersion around
the mean value of the distribution of the relative deviations (see
Fig.~\ref{fig:mass}). More relevant to the aim of this work is to
note that these different definitions of the total gravitating mass,
$M_{\rm tot}$, do not affect significantly our results on the
behaviour of the scaling relations.  We refer to a forthcoming work in
which the mass estimators applied to simulated and observed data will
be discussed in detail (see also Rasia et al. 2004).

\begin{figure}
 \epsfig{figure=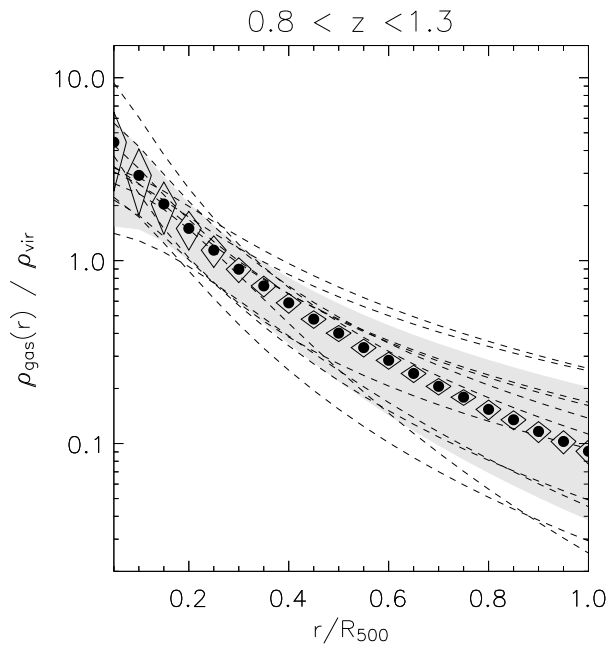,width=0.5\textwidth}
\caption{Gas density profiles at high redshift as function of
$r/R_{500}$.  {\it Dashed lines:} best-fit profiles from the \chandra
exposures of the 11 galaxy clusters with $z>0.8$ (Ettori et al. 2004).
{\it Shaded region:} $(-1, +1) \sigma$ range around the mean value of
the 11 observed profiles.  {\it Dots:} average (plotted here with the
corresponding dispersion) of the simulated density profiles at
redshift $1$ (see the right lower panel in Fig.~\ref{fig:tn_r}).  }
\label{fig:ngas}
\end{figure}

\begin{figure}
 \epsfig{figure=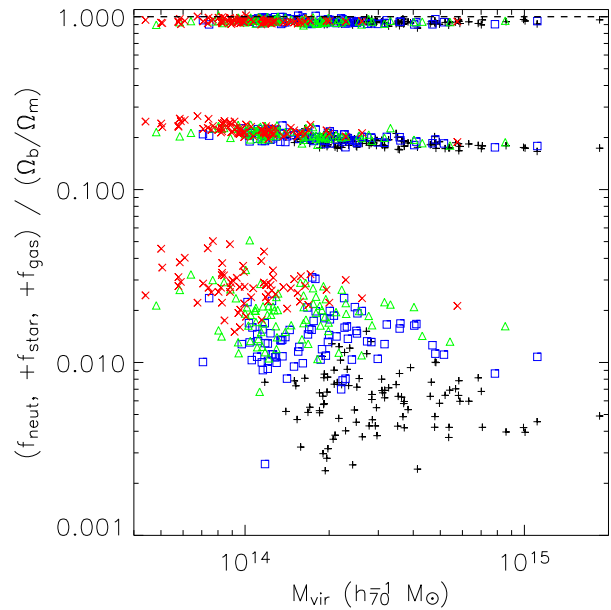,width=0.5\textwidth}
\caption{Distribution of the different components of the baryonic
budget within the cluster virial radius as a function of the total
mass.  The three concentrations of points correspond to (from the
bottom to the top) $f_{\rm neut}$ ($\sim$2 per cent), $f_{\rm neut} +
f_{\rm star}=f_{\rm cold}$ (about 20 per cent) and $f_{\rm cold} +
f_{\rm gas}$ (95 per cent) in unit of the cosmic baryon value
$\Omega_{\rm b}/\Omega_{\rm m}$.  Each component $f_i$ is equal to
$M_i / M_{\rm vir}$.  The {\it plus} symbols are for the $z=0$
objects, the {\it squares} for $z=0.5$, the {\it triangles} for
$z=0.7$ and the {\it crosses} for $z=1$.  } \label{fig:fbar}
\end{figure}

\begin{table}
\caption{Distribution of the baryons in the selected sample of
simulated galaxy clusters.  The mean values of the indicated ratios
are quoted.  } \begin{tabular}{c@{\hspace{.7em}} c@{\hspace{.7em}} c@{\hspace{.7em}}
c@{\hspace{.7em}} c@{\hspace{.7em}} c@{\hspace{.7em}} }
 \hline \\ 
 $z$ & n.obj & $\frac{M_{\rm neut}}{M_{\rm bar}}$ & $\frac{M_{\rm star}}{M_{\rm bar}}$ & $\frac{M_{\rm gas}}{M_{\rm bar}}$ & $\frac{M_{\rm bar}}{(\Omega_{\rm b}/(\Omega_{\rm m})}$ \\ \\
 $0$ & 97 & 0.007 & 0.194 & 0.799 & 0.930 \\
 $0.5$ & 83 & 0.015 & 0.198 & 0.787 & 0.944 \\
 $0.7$ & 81 & 0.021 & 0.201 & 0.779 & 0.949 \\
 $1$ & 72 & 0.031 & 0.207 & 0.761 & 0.944 \\
 \hline \\ 
\end{tabular}

\label{tab:fbar}
\end{table}

According to the multi-phase model, each gas particle of sufficiently
high overdensity is assumed to be composed of a hot ionized phase
and of a cold neutral phase, whose relative amounts depend on the
local conditions of density and temperature (Springel \& Hernquist
2003). By its nature, the neutral cold component is assumed not to
emit any X--rays.  
%SB.
%In addition, we also exclude those particles from
%the computation of X--ray emissivity having temperature below
%$3\times 10^4$ K and gas density $>500\bar\rho_{\rm bar}$, being
%$\bar\rho_{\rm bar}$ the mean baryon density. Particles inside
%clusters at such low temperature are usually at very high
%density. As such, they would provide a significant, but spurious,
%contribution to the X--ray luminosity in central cluster regions if
%they were X--ray emitting. These particles occupy a region in the
%$\rho$--$T$ plane where in principle only gas should lie that has
%already cooled, and so this gas should not be included in the
%computation of the X--ray emission (Croft et al. 2001). 
%
Following Croft et al. (2001), we also exclude those particles from
the computation of X--ray emissivity having temperature below
$3\times 10^4$ K and gas density $>500\bar\rho_{\rm bar}$, being
$\bar\rho_{\rm bar}$ the mean baryon density. 
In the following, we refer to {\em cold} gas as the gas which does not
contribute to the X--ray emission and we distinguish it between a gas
in a cold neutral phase, {\em neut}, and {\em stars}.

The X--ray luminosity of each cluster in a given energy band is
defined as
\begin{equation}
L_X\,=\,(\mu m_p)^{-2}\sum_i^{N_{\rm gas}} m_{h,i}\rho_{h,i}
\Lambda(E_1,E_2;T_i),
\label{eq:lx}
\end{equation} 
where $\Lambda(E_1,E_2;T)$ is the cooling function in the energy band
$[E_1,E_2]$. In this equation the sum runs over all $N_{\rm gas}$ gas
particles falling within the cluster selected region, and $\mu$ is the
mean molecular weight ($=0.6$ for a gas of primordial composition),
$m_p$ is the proton mass, $m_{h,i}$ and $\rho_{h,i}$ are the mass and
the density associated with the hot phase of the $i$--th gas particle,
respectively. The cooling function is computed from a Raymond--Smith
code (Raymond \& Smith 1977) by assuming zero metallicity.  
%SB
For the purpose of comparing with observational results on the
  scaling relations, in the following luminosities are always given in the
  bolometric band.
We define the emission--weighted temperature, $T_{\rm ew}$, as
\begin{equation}
T_{\rm ew}\,=\,{\sum_i^{N_{\rm gas}} m_{h,i}\rho_{h,i}
\Lambda(E_1,E_2;T_i)\,T_i\over \sum_i^{N_{\rm gas}} m_{h,i}\rho_{h,i}
\Lambda(E_1,E_2;T_i)}\,.
\label{eq:tew}
\end{equation}
We take $E_1=0.5(1+z)$ keV and $E_2=10(1+z)$ keV so as to reproduce
the observational procedure in the estimate of the temperature from
the measured photon spectrum, whose reconstruction at low energies,
say below 0.5 keV, is made hard by instrumental limitations.  
Notice however that, using simulated Chandra observations of galaxy clusters
obtained with the software package X-MAS (X-ray MAp Simulator),  Gardini
et al. (2004) showed that the emission-weighted temperature inferred
from hydrodynamical simulations can be significantly higher than the
spectroscopic value extracted from observations (see also Mathiesen 
\& Evrard 2001; Mazzotta et al. 2004). 

\begin{figure*}
\hbox{ 
 \epsfig{figure=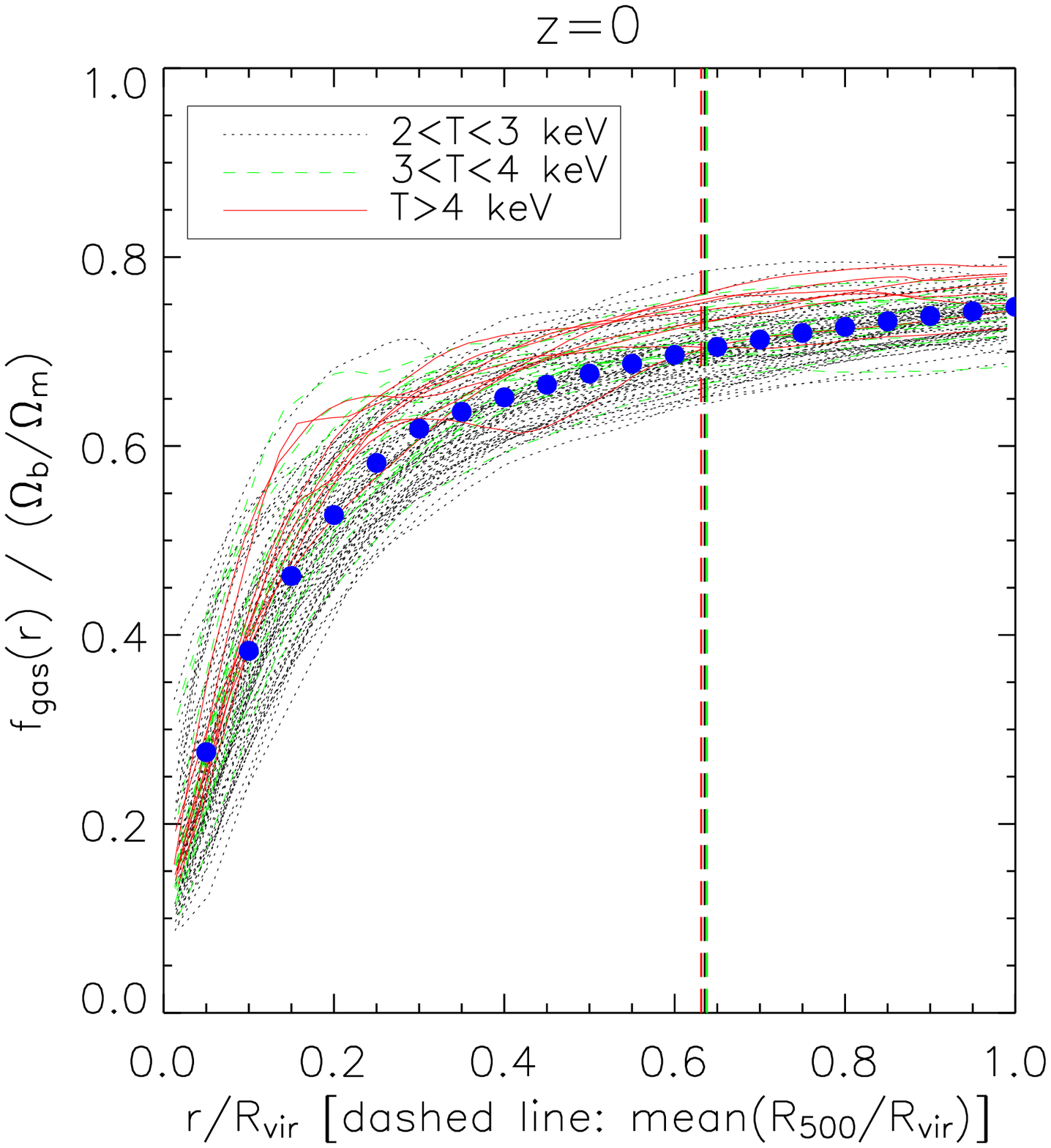,width=0.5\textwidth}
 \epsfig{figure=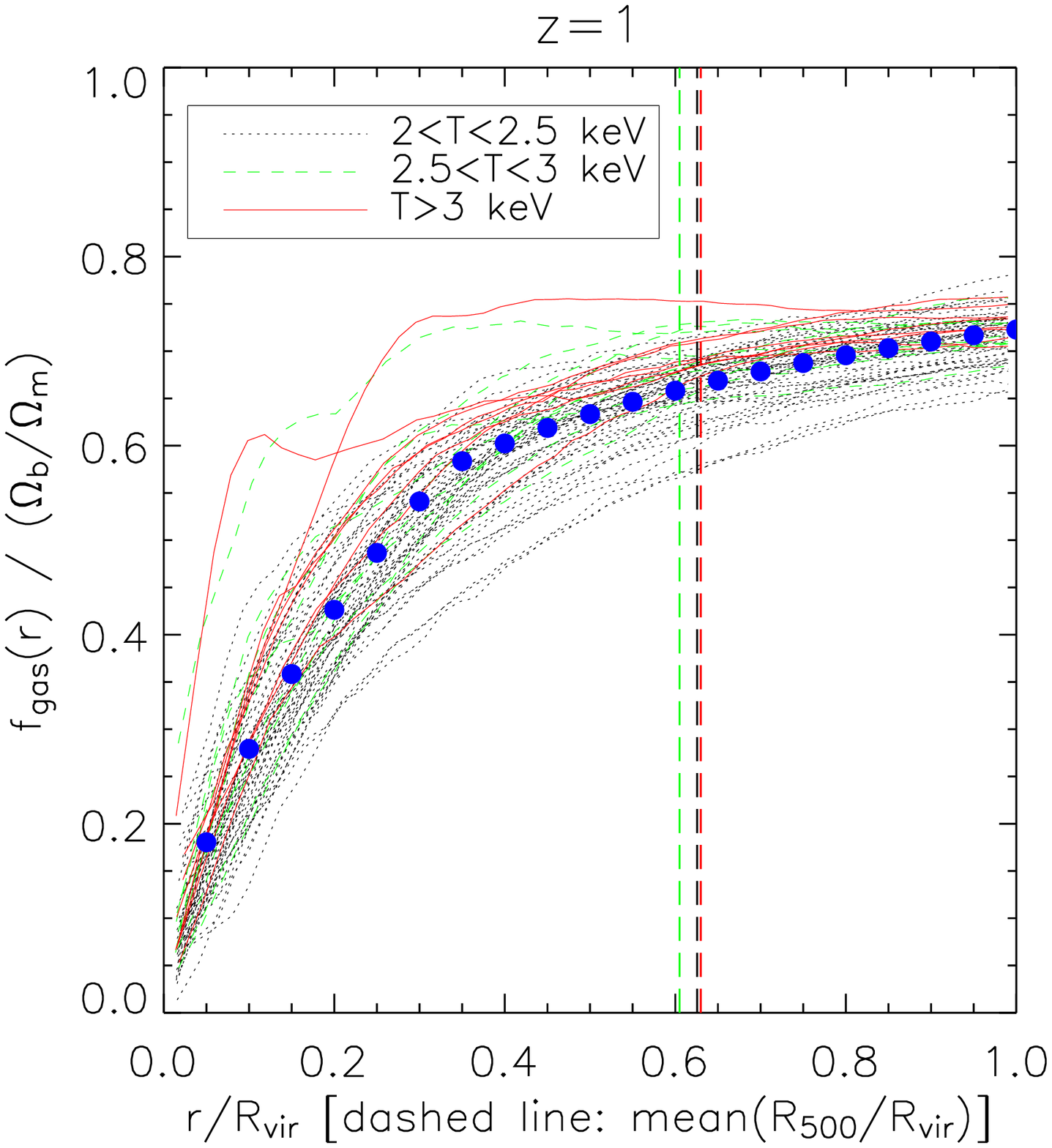,width=0.5\textwidth} } 
\caption{The gas fraction as a function of the radius in simulated clusters at
$z=0$ (left) and $z=1$ (right).
The {\it dots} represent the best-fit obtained by using the functional form
in equation~\ref{eq:fgas_r}.
} \label{fig:fgas_r}
\end{figure*}

Finally, we define the entropy of the $i$--th gas particle as
$s_i\,=\,{T_i/n_{i,e}}^{2/3}$ where $n_{i,e}$ is the number density of
free electrons associated to that gas particle.

Out of the ensemble of simulated bound structures, we analyse the 97,
83, 81 and 72 systems at redshift 0, 0.5, 0.7 and 1, respectively,
that have an emission weighted gas temperature, $T_{\rm ew}$, 
within $\Delta=500$ larger than 2 keV. This selection allows to consider
only the objects where the gravitational collapse dominates the
energetic budget and that present a local temperature function and
luminosity--temperature relation in good agreement with the observed
ones (see Paper I). 

In Figure~\ref{fig:tn_r}, we plot the gas temperature and density
profiles of the examined clusters at redshift $0$ and $1$.  At higher
redshift, (i) the radial profiles tend to be more spread around the
mean value, suggesting that many systems are still in formation, (ii)
the temperature distribution appears slightly flatter within the
virial radius, decreasing by about 50 per cent from $0.2 \times R_{\rm
vir}$ to $0.8 \times R_{\rm vir}$ (to be compared with a reduction by
$\sim$60 per cent at $z=0$), and (iii) the central ($r < 0.2 \times
R_{\rm vir}$) gas density is less peaked by about 50 per cent than
what observed at $z=0$.  However, the mean polytropic index $\gamma$,
measured as the coefficient of the linear fit $\log T_{\rm gas}(r) = k
+(\gamma-1) \times \log \rho_{\rm gas}(r)$, remains almost constant
with redshift, varying from 1.18 (rms: 0.04) at $z=0$ to 1.16 (rms:
0.06) at $z=1$. We conclude that the overall shape of the gas
temperature and density profiles does not change significantly with
redshift, even though local radial variations are present as a
consequence of the different dynamical state of the objects, with
dynamically younger systems being located at higher redshift.

In Figure~\ref{fig:ngas} we show the comparison between simulated and
observed gas density profiles at $0.8 < z <1.3$. The observed profiles
are recovered from the best-fit results obtained by applying a
$\beta-$model to deep \chandra exposures (Ettori et al. 2004) and are
normalized by the corresponding $\rho_{\rm vir}$ (see caption in
Fig.~\ref{fig:tn_r}) as done for simulated profiles.  The overall
shape seems in good agreement with the mean estimated values, even
though the scatter in the observed data (indicated by the gray region
in Fig.~\ref{fig:ngas}) is definitely larger than what is measured in the
simulated ones at $z=1$.  This suggests that either the real data at
high$-z$ are subjected to a more complicated dynamical history than
actually simulated or the latter ones are more regular due to the
criteria adopted in selecting the dark matter halos.  In detail, the
more significant deviations are within $0.1 \times R_{500}$ and above
$0.7 \times R_{500}$, where the steeper simulated profiles have mean
values higher by 10-20 per cent and lower by 20 per cent (by 25 per
cent at $R_{500}$) than the observed ones, respectively.  The
difference remains within few per cent in the range $(0.1-0.7) \times
R_{500}$. On the other hand, once we limit our analysis to the hottest
systems by selecting the 10 objects with the highest $T_{\rm
ew}(<R_{500})$, the discrepancy between the average profiles becomes
more noticeable: deviations larger than 10 per cent are present within
$0.4 \times R_{500}$ and beyond $0.8 \times R_{500}$.  However, once
we compare the rough estimate of the expected mean gas mass by
integrating these average profiles over the cluster volume up to
$R_{500}$, we obtain a reasonable agreement between the observed and
simulated values when only the hottest systems are analysed.  A lower
estimate by about 10 per cent is measured when the average gas density
profile is obtained from all the simulated $z=1$ objects.
 
The few observational data available on the gas temperature profiles
at high$-z$ (e.g. Jeltema et al. 2001, Arnaud et al. 2002) indicate
isothermality up to half the virial radius but with typical relative
uncertainties of 20 per cent or more (90 per cent confidence level) on
each temperature measurement of the 3-4 radial bins.  Mismatches
between observed and simulated temperature profiles of the same order
as highlighted in Paper I for local systems are probably present here
but can not be confirmed. We refer the reader to section 3.3 and 3.4
of that paper for a detailed discussion of these discrepancies.

\section{The distribution of baryons}

The distribution of the baryons, and of the hot X-ray emitting gas in
particular, is a diagnostic of the energetic phenomena that take place
during the formation and evolution of galaxy clusters. Since this
distribution affects the estimates of the gas density, entropy and
emissivity, the analysis of how baryons populate the cluster potential
well provides information on the properties of the integrated
measurements.

Overall, we find between 1 (at redshift $0$) and 3 (at $z=1$) per cent
of the baryons to be present in the cold phase, about 20 per cent in
stars and between 77 (at $z=1$) and 80 (at $z=0$) per cent in the hot,
X-ray emitting plasma.  We also note a slight decrease from 7 to 5 per
cent in the depletion of the baryons with respect to the assumed
cosmic baryonic budget from $z=0$ to 1 (see Table~\ref{tab:fbar}).
These values of depletion, that have to be considered as lower limits
considering the measured over-production of stars, are however
comparable to what predicted from adiabatic simulations such as the
ones considered in the Santa Barbara Project (Frenk et al. 1999) that
indicate a mean $f_{\rm b} / (\Omega_{\rm b}/\Omega_{\rm m})$ of 0.92
(rms: 0.06).

The overall amount of cold gas, $M_{\rm neut}+ M_{\rm star}$ is mainly
composed of stars formed in high density regions and accounts for
about 20 per cent of the total baryon budget (see average value in our
samples in Table~\ref{tab:fbar}), at variance with a $\sim 10$ per
cent estimate observed in nearby clusters (e.g. Lin et al.  2003).
%SB. 
We point out that the major part of the star overproduction in
simulated clusters takes place in the central cD galaxy, where the
action of the galactic winds is not efficient enough to prevent
overcooling. The lack of observational evidence for low--temperature
gas in central cluster regions (e.g., Peterson et al. 2003) calls for
the need of a central heating mechanism. We argue that this mechanism,
which is not included in the present simulations, should reduce
the resulting amount of stars.
Moreover, there is a mild evidence for a higher efficiency to form
stars in smaller systems (a trend also found in observations, see
Fig.~7 in Lin et al. 2003) and at higher redshift
(Fig.~\ref{fig:fbar}). As Springel \& Hernquist (2003) suggest, this
is an evidence for a less efficient formation of cooling flows in
halos with virial temperature above $10^7$ K.  However, an analysis of
our simulated clusters (Murante et al. 2004) shows that we have a
diffuse star component, not associated with any galaxy, whose mass
ranges from 10 per cent to more than 40 per cent of the total stellar
mass of the cluster, being higher for more massive objects.  Thus, the
disagreement between the amount of cold baryons present in simulations
and those observed could be lessened. Furthermore, depending on the
spatial distribution of the diffuse component, the estimates of the
star formation efficiencies could be slightly biased both in
observations and in simulations, in the direction of having a smaller
efficiency for massive clusters.

The X-ray emitting gas fraction, $f_{\rm gas}(<r) = M_{\rm gas}(<r) /
M_{\rm tot}(<r)$, increases when moving outward to $(0.3-0.4) \times
R_{\rm vir}$ and flattens then toward a constant value that is about
$0.8 \times f_{\rm bar} \approx 0.75 \times \Omega_{\rm b}/\Omega_{\rm
m}$ (Fig.~\ref{fig:fgas_r}).  This confirms that the hot ICM is more
extended than the dark matter distribution, as already known from
observational data (e.g. David et al. 1995, Ettori \& Fabian 1999),
and that one needs to sample the cluster X-ray emission out to $\sim
0.4 \times R_{\rm vir} \approx 0.7 \times R_{500}$ to fully recover
the baryon fraction in the form of hot plasma.

\begin{figure}
% \hbox{
 \epsfig{figure=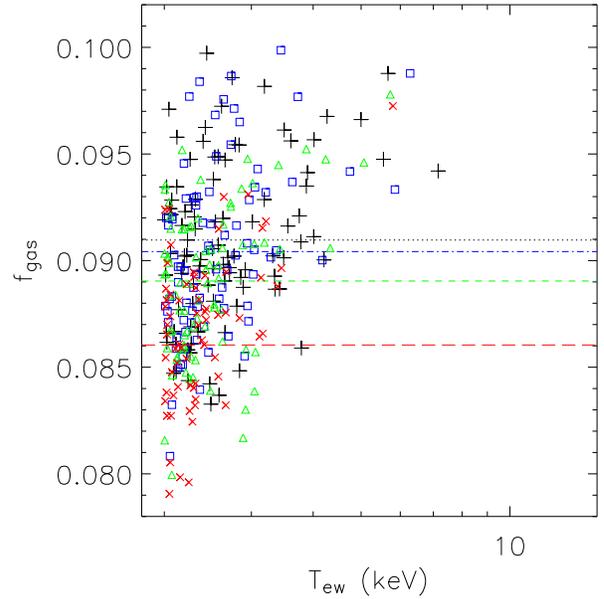,width=0.5\textwidth}
\caption{The gas fraction within $\Delta=500$ as a function of the gas
temperature.
% (left) and total mass (right).  
The symbols of the
points are as in caption of Fig.~\ref{fig:fbar}.  The horizontal lines
indicate the mean values of the gas fraction at different redshift ({\it
dotted line}: $z=0$; {\it dash-dotted line}: $z=0.5$; {\it dashed
line}: $z=0.7$; {\it long dashes}: $z=1$).  } \label{fig:fgas_tm}
\end{figure}

We find that a simple functional form can reproduce this dependence
upon the radius quite well in the regions beyond the inner steep
gradient, $f_{\rm gas}(<r) \propto (2 r)^{\eta}/(R_{\rm
vir}+r)^{\eta}$, with $\eta$ about 0.2--0.3, whereas a further
suppression is required to reproduce the profile within $\sim 0.3
\times R_{\rm vir}$:
\begin{equation}
\frac{f_{\rm gas}(<r)}{f_{\rm gas}(<R_{\rm vir})} = 
\min \left( \frac{r}{r_{\rm s}}, 1 \right)^{\eta} \times 
\left( \frac{2 \ r/R_{\rm vir}}{1+ r/R_{\rm vir}} \right)^{\eta}, 
\label{eq:fgas_r}
\end{equation} 
with $r_{\rm s} \approx 0.3 R_{\rm vir}$.
The best-fit results obtained by applying this functional form to the
average radial profile of $f_{\rm gas}$ are: $f_{\rm gas}(<R_{\rm vir})
= 0.75 \times \left( \Omega_{\rm b}/\Omega_{\rm m} \right)$, $\eta$
between 0.24 and 0.32, and $r_{\rm s}$ increasing from $0.28R_{\rm
vir}$ to $0.34R_{\rm vir}$ when systems at higher redshift are
considered.

To explore the gas mass fraction as a function of radius in the
hottest systems that are more accessible to the observational
analysis, we select the 10 hottest systems at redshift $0$ and $1$
(average value of $T_{\rm ew}(<R_{500}) =$ 4.8 and 3.4 keV,
respectively). We measure then a mean (standard deviation) $f_{\rm
gas}/(\Omega_{\rm b}/\Omega_{\rm m})$ of 0.638 (0.030) and 0.567
(0.071) at $R_{2500} \approx 0.28 R_{\rm vir}$, 0.732 (0.020) and
0.698 (0.026) at $R_{500}$, 0.760 (0.021) and 0.731 (0.015) within
$R_{\rm vir}$.  Even though a larger contribution of hot gas to the
cosmic fraction is measured at higher redshift, these measurements are
in reasonable agreement once the associated scatter is considered.

\begin{table*}
\caption{Best-fit results on the local ($z=0$) scaling relations 
(dashed and dotted lines in the plots; see equation~\ref{eq:fit})
of simulated galaxy clusters.
The temperature, $T$, is in unit of 6 keV;
the bolometric luminosity, $L$, in $10^{44}$ erg s$^{-1}$;
the total mass, $M_{\rm tot}$, in $10^{14} M_{\odot}$;
the gas mass, $M_{\rm gas}$, in $10^{13} M_{\odot}$. All these
quantities are estimated within $R_{500}$. The entropy, $S$, is in unit 
of keV cm$^{-2}$ and is measured at $0.1 \times R_{200}$. 
When the slope $A$ is fixed, we estimate the error-weighted mean
of $(\log Y -A \log X)$ and evaluate the error after resampling $Y$ and $X$
by 1\,000 times according to their uncertainties.
% The scatter on $Y$ is measured as $\left[ \sum_{j=1,N}   
% \left(\log Y_j -\alpha -A \log X_j \right)^2 /N \right]^{1/2}$.
% Note that the scatter along the X-axis can be estimated as
% $\sigma_{\log X} = \sigma_{\log Y} / A$.
The best-fit on the evolution parameter $B$ (see equation~\ref{eq:bfit})
is obtained by considering (i) all the objects at redshift $\ge 0.5$,
(ii) those at $\ge 0.7$ and (iii) only the clusters at $z=1$.
The quoted errors are at $1 \sigma$ level (68.3 per cent level of confidence 
and $\Delta \chi^2=1$ for one interesting parameter).
} \begin{tabular}{l@{\hspace{.7em}} c@{\hspace{.7em}} c@{\hspace{.7em}} c@{\hspace{.7em}}
c@{\hspace{.7em}} c@{\hspace{.7em}} c@{\hspace{.7em}} }
 \hline \\ 
 relation $(Y-X)$ & $\alpha$ & $A$ & & $B$ ($z \ge 0.5$) & $B$ ($z \ge 0.7$) & $B$ ($z=1$) \\ 
  & & & & & &  \\ 
 \hline \\ 
 $F_z M_{\rm tot} - T$ & $1.07 (\pm0.06)$ & $2.08 (\pm0.15)$ & & $-0.12 (\pm0.05)$ & $-0.14 (\pm0.05)$ & $-0.20 (\pm0.07)$ \\ 
 & $0.87 (\pm0.01)$ & $1.50$ (fixed) & & $-0.20 (\pm0.04)$ & $-0.22 (\pm0.04)$ & $-0.30 (\pm0.05)$ \\ 
  & &  & \\ 
 $F_z M_{\rm gas} - T$ & $1.14 (\pm0.08)$ & $2.40 (\pm0.20)$ & & $-0.12 (\pm0.05)$ & $-0.14 (\pm0.06)$ & $-0.22 (\pm0.08)$ \\ 
 & $0.83 (\pm0.01)$ & $1.50$ (fixed) & & $-0.26 (\pm0.03)$ & $-0.30 (\pm0.03)$ & $-0.38 (\pm0.04)$ \\ 
  & &  & \\ 
 $F_z^{-1} L - T$ & $1.23 (\pm0.14)$ & $3.33 (\pm0.35)$ & & $-0.76 (\pm0.08)$ & $-0.76 (\pm0.09)$ & $-0.80 (\pm0.11)$ \\ 
 & $0.76 (\pm0.01)$ & $2.00$ (fixed) & & $-0.96 (\pm0.04)$ & $-0.98 (\pm0.04)$ & $-1.04 (\pm0.05)$ \\ 
  & &  & \\ 
 $F_z^{-1} L - F_z M_{\rm tot}$ & $-0.47 (\pm0.04)$ & $1.57 (\pm0.11)$ & & $-0.56 (\pm0.05)$ & $-0.56 (\pm0.05)$ & $-0.50 (\pm0.07)$ \\ 
 & $-0.39 (\pm0.02)$ & $1.33$ (fixed) & & $-0.68 (\pm0.04)$ & $-0.66 (\pm0.04)$ & $-0.64 (\pm0.05)$ \\ 
  & &  & \\ 
 $F_z^{-1} L - F_z M_{\rm gas}$ & $-0.35 (\pm0.03)$ & $1.38 (\pm0.07)$ & & $-0.58 (\pm0.03)$ & $-0.56 (\pm0.03)$ & $-0.50 (\pm0.04)$ \\ 
 & $-0.34 (\pm0.01)$ & $1.33$ (fixed) & & $-0.60 (\pm0.02)$ & $-0.58 (\pm0.03)$ & $-0.54 (\pm0.04)$ \\ 
  & &  & \\ 
 $f_{\rm gas} - T$ & $-1.01 (\pm0.01)$ & $0.09 (\pm0.02)$ & & $-0.04 (\pm0.03)$ & $-0.06 (\pm0.04)$ & $-0.08 (\pm0.05)$ \\ 
 & $-1.04 (\pm0.01)$ & $0.00$ (fixed) & & $-0.06 (\pm0.03)$ & $-0.06 (\pm0.04)$ & $-0.08 (\pm0.04)$ \\ 
  & &  & \\ 
 $F_z^{4/3} S - T$ &$2.81 (\pm0.06)$ & $1.42 (\pm0.17)$ & & $0.26 (\pm0.04)$ & $0.24 (\pm0.04)$ & $0.18 (\pm0.05)$ \\
 & $2.66 (\pm0.01)$ & $1.00$ (fixed) & & $0.22 (\pm0.02)$ & $0.18 (\pm0.02)$ & $0.12 (\pm0.03)$ \\ 
 \hline \\ 
\end{tabular}

\label{tab:fit}
\end{table*}

\begin{table*}
\caption{Best-fit results on the simulated and observed high$-z$ ($z \ge 0.5$) samples.
The relations and the parameters are defined in Table~\ref{tab:fit}.
The evolution parameter $B$ is estimated by using the local
relations indicated in Table~\ref{tab:fit} for simulated objects,
and the best-fit values in Finoguenov et al. (2001, $M_{\rm tot}-T$),
Mohr et al. (1999, $M_{\rm gas}-T$ and $f_{\rm gas}-T$), Ettori et al. (2002, $L-T$),
Reiprich \& B\"ohringer (2002, $L-M_{\rm tot}$) and Ponman et al. (2003,
$S-T$) for the observed 22 clusters with $z \ge 0.5$ 
(Ettori et al. 2004). The deviations in $\sigma$ between the parameter $B$ 
as estimated in the simulations and observations are indicated in the column
``$\sigma_B$''.}
\begin{tabular}{l@{\hspace{.7em}} c@{\hspace{.7em}} c@{\hspace{.7em}} c@{\hspace{.7em}}
c@{\hspace{.7em}} c@{\hspace{.7em}} c@{\hspace{.7em}} c@{\hspace{.7em}} c@{\hspace{.7em}} c@{\hspace{.7em}} }
 \hline \\ 
 relation $(Y-X)$ & $\alpha$ & $A$ & $B$ & & {\rm $\alpha$} & {\rm $A$} & {\rm $B$} & & $\sigma_B$ \\ 
  & \multicolumn{3}{c}{{\rm from simulations}} & & \multicolumn{3}{c}{{\rm from observations}} \\ 
 \hline \\ 
 $F_z M_{\rm tot} - T$ & $1.29 (\pm0.12)$ & $2.72 (\pm0.31)$ & $-0.12 (\pm0.05)$ & & $0.85 (\pm0.04)$ & $1.89 (\pm0.45)$ & $0.24 (\pm0.25)$ & & $-1.44$ \\
 & $0.82 (\pm0.01)$ & $1.50$ (fixed) & $-0.20 (\pm0.04)$ & & $0.88 (\pm0.02)$ & $1.50$ (fixed) \\
  & &  & \\ 
 $F_z M_{\rm gas} - T$ & $1.49 (\pm0.18)$ & $3.35 (\pm0.46)$ & $-0.12 (\pm0.05)$ & & $0.83 (\pm0.04)$ & $2.63 (\pm0.36)$ & $-0.52 (\pm0.28)$ & & $1.42$ \\
 & $0.83 (\pm0.01)$ & $1.50$ (fixed) & $-0.26 (\pm0.03)$ & & $0.98 (\pm0.02)$ & $1.50$ (fixed) \\
  & &  & \\ 
 $F_z^{-1} L - T$ & $1.45 (\pm0.22)$ & $4.33 (\pm0.54)$ & $-0.76 (\pm0.08)$ & & $0.58 (\pm0.07)$ & $4.12 (\pm0.75)$ & $-2.20 (\pm0.46)$ & & $3.12$ \\
 & $0.55 (\pm0.01)$ & $2.00$ (fixed) & $-0.96 (\pm0.04)$ & & $0.75 (\pm0.03)$ & $2.00$ (fixed) \\
  & &  & \\ 
 $F_z^{-1} L - F_z M_{\rm tot}$ & $-0.61 (\pm0.02)$ & $1.57 (\pm0.07)$ & $-0.56 (\pm0.05)$ & & $-1.33 (\pm0.60)$ & $2.24 (\pm0.71)$ & $-2.28 (\pm0.31)$ & & $5.44$ \\
 & $-0.55 (\pm0.01)$ & $1.33$ (fixed) & $-0.68 (\pm0.04)$ & & $-0.38 (\pm0.04)$ & $1.33$ (fixed) \\
  & &  & \\ 
 $F_z^{-1} L - F_z M_{\rm gas}$ & $-0.48 (\pm0.01)$ & $1.31 (\pm0.04)$ & $-0.58 (\pm0.03)$ & & $-0.70 (\pm0.06)$ & $1.55 (\pm0.07)$ & $-$ \\
 & $-0.48 (\pm0.01)$ & $1.33$ (fixed) & $-0.60 (\pm0.02)$ & & $-0.49 (\pm0.02)$ & $1.33$ (fixed) \\
  & &  & \\ 
 $f_{\rm gas} - T$ & $-0.96 (\pm0.03)$ & $0.24 (\pm0.09)$ & $-0.04 (\pm0.03)$ & & $-1.03 (\pm0.04)$ & $0.60 (\pm0.49)$ & $-0.64 (\pm0.22)$ & & $2.75$ \\
 & $-1.05 (\pm0.01)$ & $0.00$ (fixed) & $-0.06 (\pm0.03)$ & & $-0.89 (\pm0.02)$ & $0.00$ (fixed) \\
  & &  & \\ 
 $F_z^{4/3} S - T$ &$3.19 (\pm0.17)$ & $2.22 (\pm0.42)$ & $0.26 (\pm0.04)$ & & $2.75 (\pm0.02)$ & $0.70 (\pm0.23)$ & $0.68 (\pm0.06)$ & & $-5.87$ \\
 & $2.71 (\pm0.01)$ & $1.00$ (fixed) & $0.22 (\pm0.02)$ & & $2.65 (\pm0.01)$ & $1.00$ (fixed) \\
 \hline \\ 
\end{tabular}

\label{tab:fit2}
\end{table*}

\begin{figure}
% \hbox{
 \epsfig{figure=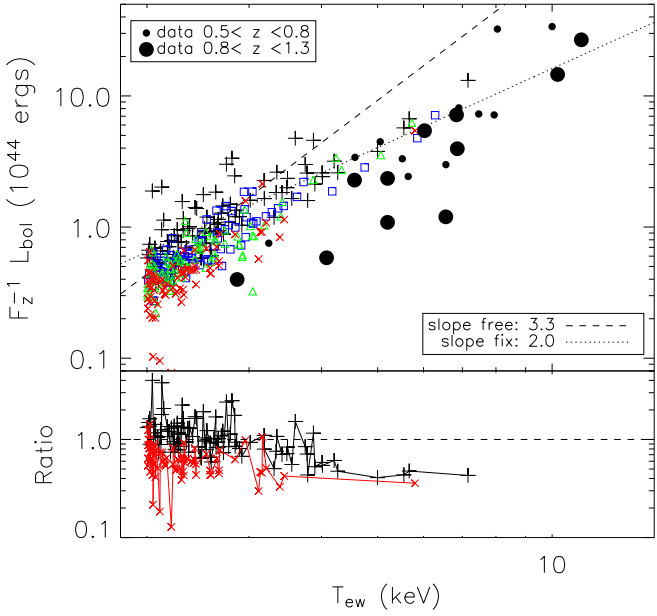,width=0.5\textwidth}
\caption{The local $L-T$ relation estimated at $\Delta=500$ as
best-fit from the simulated clusters at $z=0$ ({\it plus
symbols}). Dotted line: slope fixed to the predicted value. Dashed
line: slope free.  The simulated objects at $z=0.5$ ({\it squares}),
$z=0.7$ ({\it triangles}) and $z=1$ ({\it crosses}) are compared to
the observed dataset from Ettori et al. (2004) with $0.5 < z <0.8$
({\it small dots}) and $z>0.8$ ({\it large dots}).  {\bf (Lower
panel)} Ratio between the measured luminosities at $z=0$ ({\it plus
symbols}) and $z=1$ ({\it crosses}) and the best-fit local power-law.
% {\it (Right panel)} Constraints on the evolution parameter $B$ when all
% the $z \ge 0.5$ objects are considered.
}
\label{fig:lt}
\end{figure}

\begin{figure}
% \hbox{
 \epsfig{figure=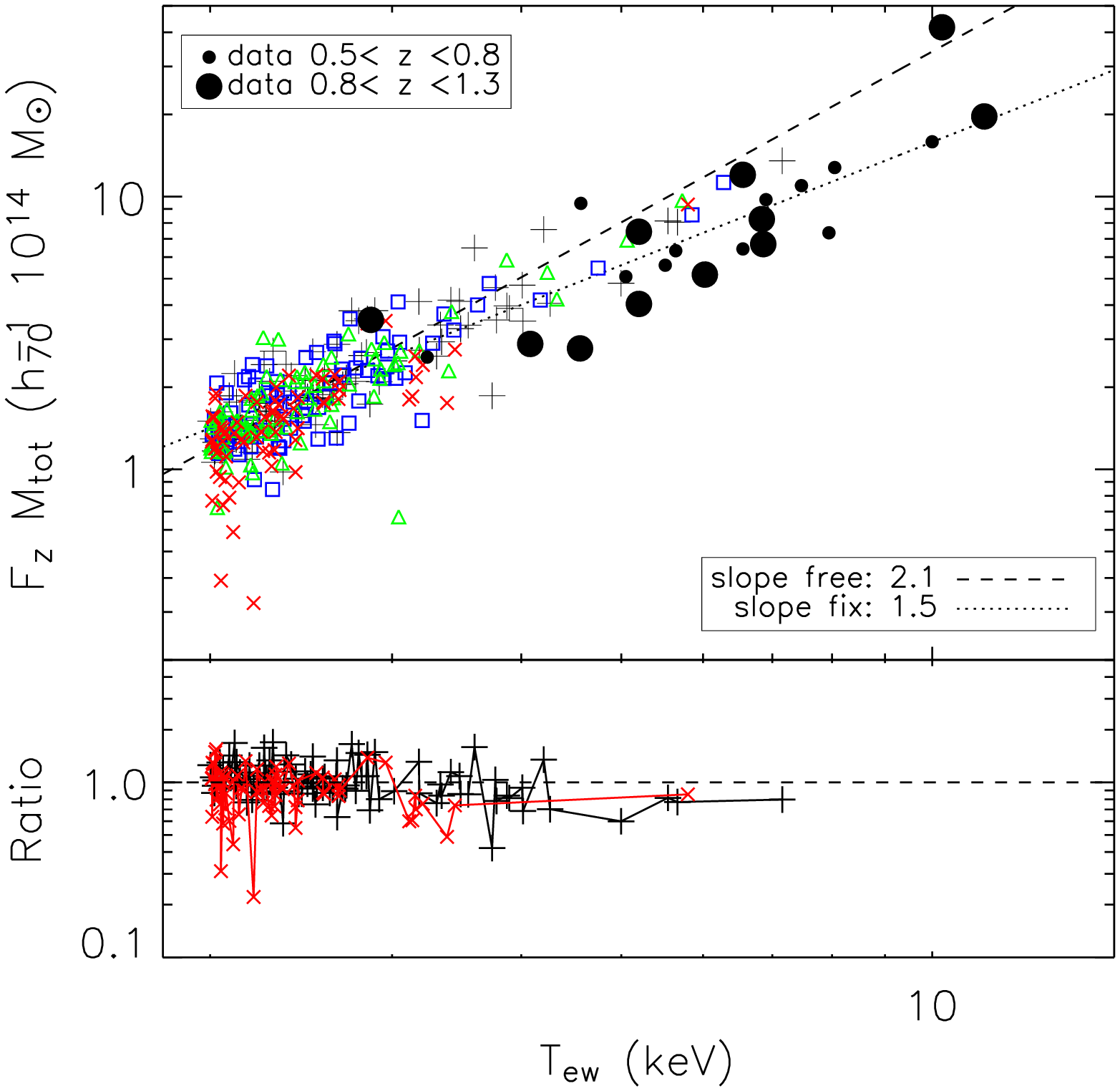,width=0.5\textwidth}
\caption{The same as in Fig.\ref{fig:lt}, but for the $M_{\rm tot}-T$
 relation.
%Local $M_{\rm tot}-T$ relation estimated as best-fit from the 
%97 simulated clusters at $z=0$ ({\it black plus}). 
%Dotted line: slope fixed to the predicted value. Dashed line: slope free.
%The simulated objects at $z=0.5$ ({\it blue squares}),
%$z=0.7$ ({\it green triangles}) and $z=1$ ({\it red crosses}) are compared
%to the observed dataset from Ettori et al. (2004) with $0.5 < z <0.8$ 
%({\it small black dots}) and $z>0.8$ ({\it large black dots}).
%
% {\it (Right panel)} Constraints on the evolution parameter $B$ when all
% the $z \ge 0.5$ objects are considered.
}
\label{fig:mt}
\end{figure}

\begin{figure}
% \hbox{
 \epsfig{figure=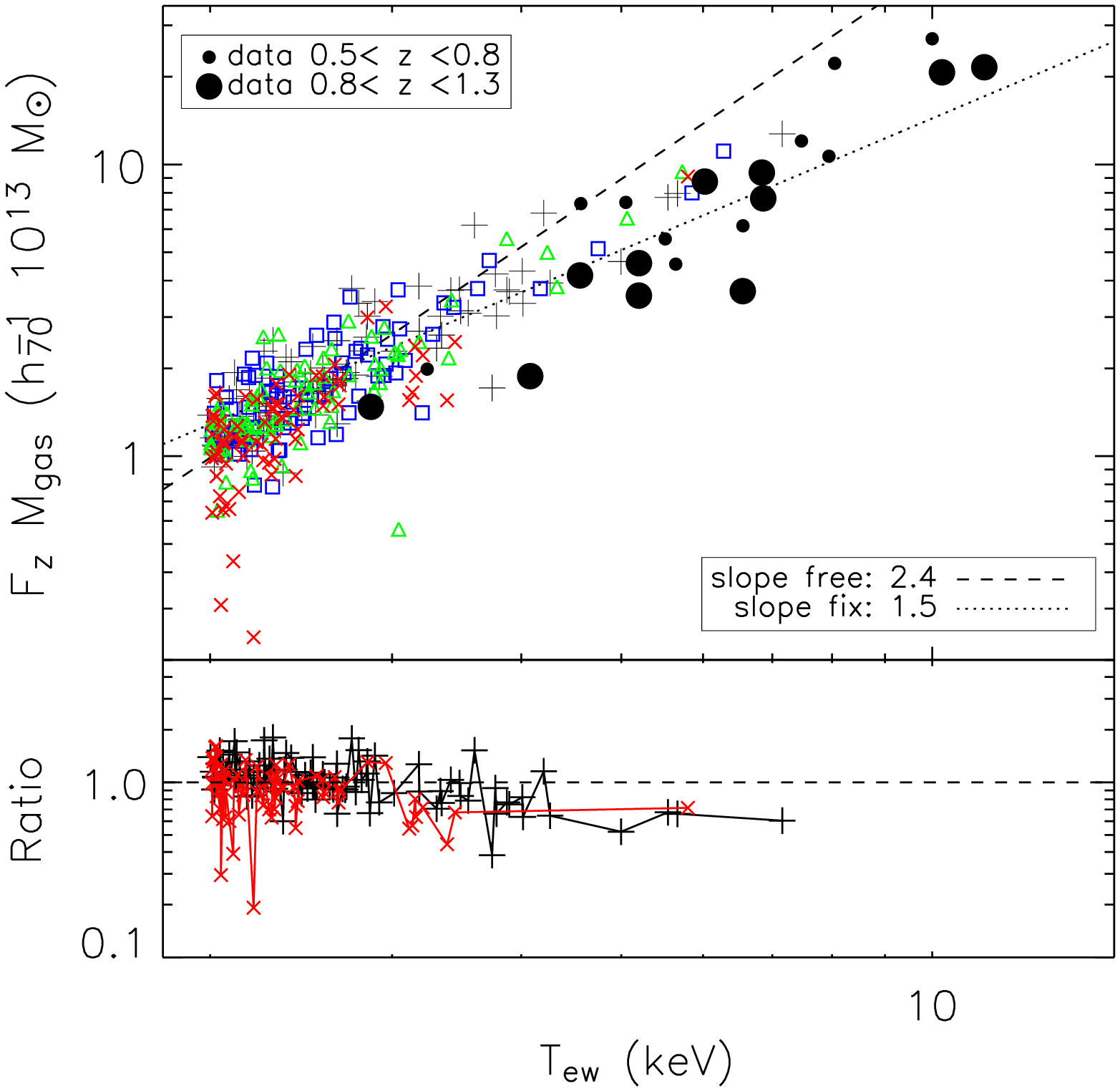,width=0.5\textwidth}
\caption{The same as in Fig.\ref{fig:lt}, but for the $M_{\rm gas}-T$ 
 relation. 
%Local $M_{\rm gas}-T$ relation estimated as best-fit from the 
%97 simulated clusters at $z=0$ ({\it black plus}). 
%Dotted line: slope fixed to the predicted value. Dashed line: slope free.
%The simulated objects at $z=0.5$ ({\it blue squares}),
%$z=0.7$ ({\it green triangles}) and $z=1$ ({\it red crosses}) are compared
%to the observed dataset from Ettori et al. (2004) with $0.5 < z <0.8$
%({\it small black dots}) and $z>0.8$ ({\it large black dots}).
%
% {\it (Right panel)} Constraints on the evolution parameter $B$ when all
% the $z \ge 0.5$ objects are considered.
}
\label{fig:mgt}
\end{figure}

\begin{figure}
% \hbox{
 \epsfig{figure=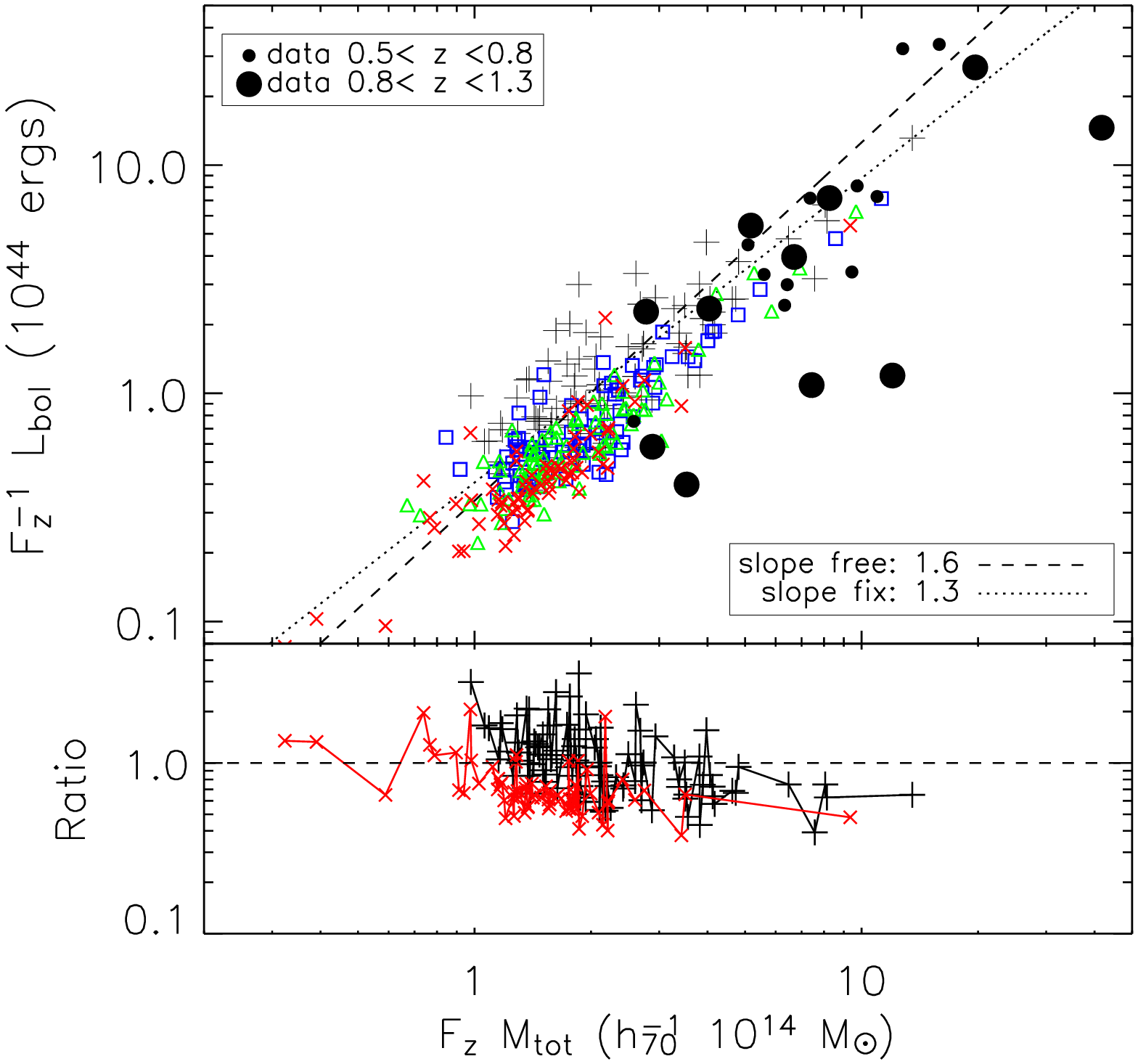,width=0.5\textwidth}
\caption{The same as in Fig.\ref{fig:lt}, but for the $L-M_{\rm tot}$  
 relation.
%Local $L-M_{\rm tot}$ relation estimated as best-fit from the 
%97 simulated clusters at $z=0$ ({\it black plus}). 
%Dotted line: slope fixed to the predicted value. Dashed line: slope free.
%The simulated objects at $z=0.5$ ({\it blue squares}),
%$z=0.7$ ({\it green triangles}) and $z=1$ ({\it red crosses}) are compared
%to the observed dataset from Ettori et al. (2004) with $0.5 < z <0.8$
%({\it small black dots}) and $z>0.8$ ({\it large black dots}).
%
% {\it (Right panel)} Constraints on the evolution parameter $B$ when all
% the $z \ge 0.5$ objects are considered.
}
\label{fig:lm}
\end{figure}

\begin{figure}
% \hbox{
 \epsfig{figure=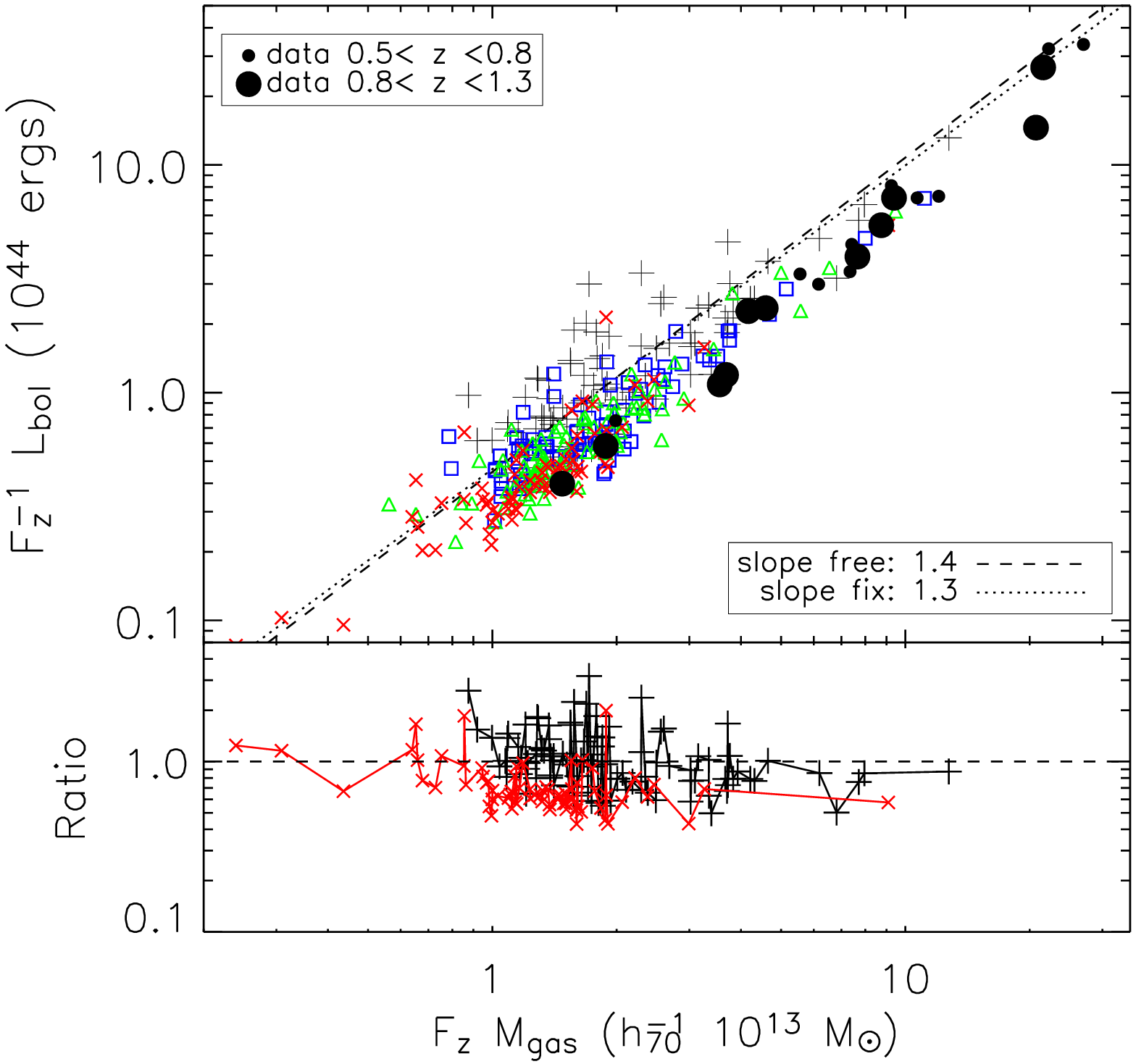,width=0.5\textwidth}
\caption{The same as in Fig.\ref{fig:lt}, but for the $L-M_{\rm gas}$   
 relation.
%Local $L-M_{\rm gas}$ relation estimated as best-fit from the 
%97 simulated clusters at $z=0$ ({\it black plus}). 
%Dotted line: slope fixed to the predicted value. Dashed line: slope free.
%The simulated objects at $z=0.5$ ({\it blue squares}),
%$z=0.7$ ({\it green triangles}) and $z=1$ ({\it red crosses}) are compared
%to the observed dataset from Ettori et al. (2004) with $0.5 < z <0.8$
%({\it small black dots}) and $z>0.8$ ({\it large black dots}).
%
% {\it (Right panel)} Constraints on the evolution parameter $B$ when all
% the $z \ge 0.5$ objects are considered.
}
\label{fig:lmg}
\end{figure}

\begin{figure}
% \hbox{
 \epsfig{figure=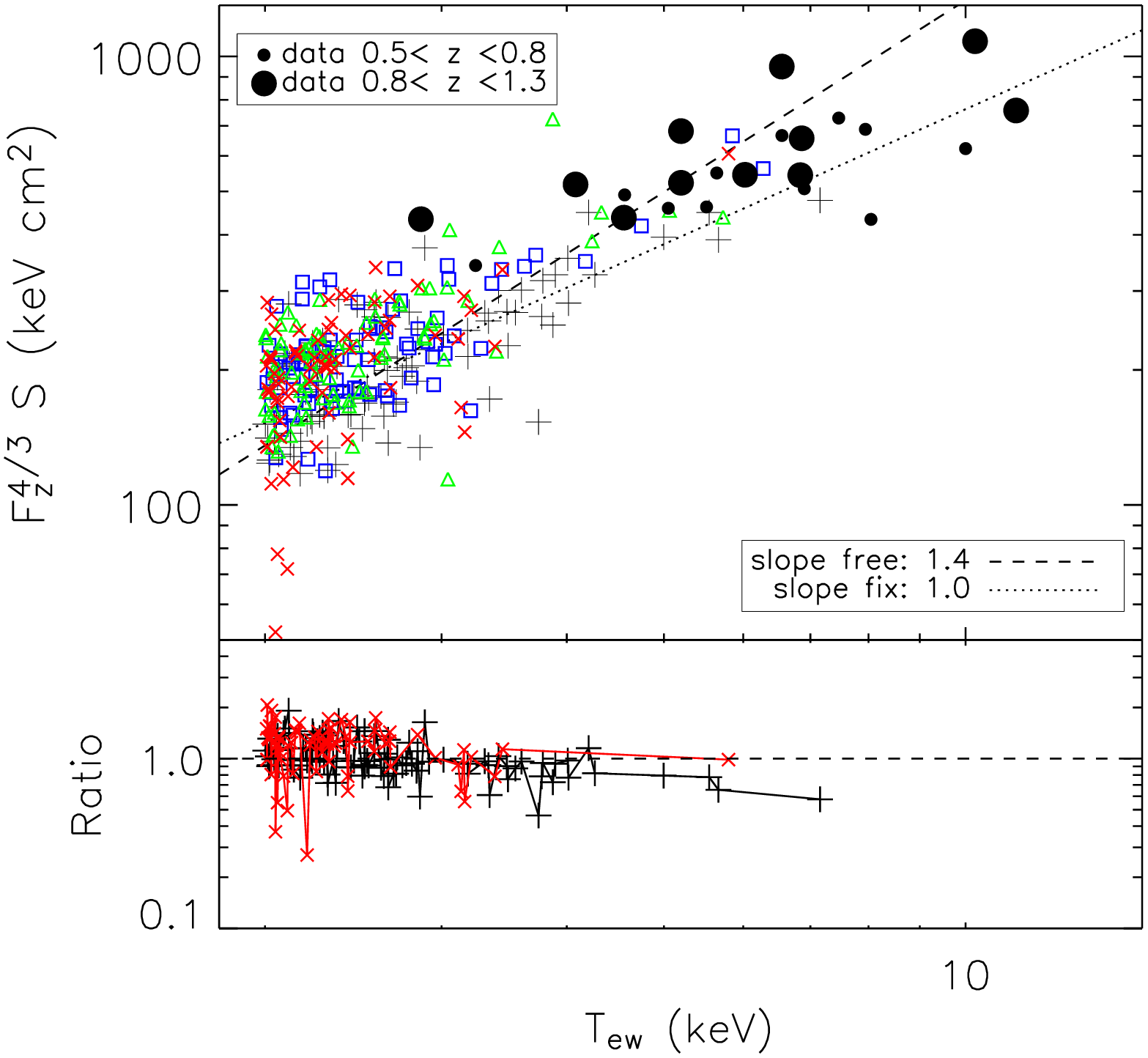,width=0.5\textwidth}
\caption{The same as in Fig.\ref{fig:lt}, but for the $S-T$   
 relation evaluated at $R_{0.1} = 0.1 \times R_{200}$.
%Local $S-T$ relation estimated as best-fit from the 
%97 simulated clusters at $z=0$ ({\it black plus}). 
%Dotted line: slope fixed to the predicted value. Dashed line: slope free.
%The simulated objects at $z=0.5$ ({\it blue squares}),
%$z=0.7$ ({\it green triangles}) and $z=1$ ({\it red crosses}) are compared
%to the observed dataset from Ettori et al. (2004) with $0.5 < z <0.8$
%({\it small black dots}) and $z>0.8$ ({\it large black dots}).
%
% {\it (Right panel)} Constraints on the evolution parameter $B$ when all
% the $z \ge 0.5$ objects are considered.
}
\label{fig:st}
\end{figure}

\section{The scaling relations and their evolution}

Under the assumptions that the smooth and spherically symmetric intra-cluster
medium (ICM) emits by bremsstrahlung and is in hydrostatic
equilibrium with the underlying dark matter potential, one
can relate the bolometric luminosity, $L$,
gas temperature, $T$, gas entropy, $S$, gas mass, $M_{\rm gas}$, 
and total mass, $M_{\rm tot}$ in a simple and straightforward way.
For instance, the equation of hydrostatic equilibrium, 
$d (\rho_{\rm gas} T)
/dr \approx \rho_{\rm gas} G M_{\rm tot} / r^2$, allows to write
$M_{\rm tot} \propto T R$ that, once combined with 
the definition of the total mass within a given overdensity $\Delta_z$,
$M_{\rm tot} \propto E_z^2 \Delta_z R^3$, implies $E_z \Delta_z^{1/2} M_{\rm tot} 
\propto T^{3/2}$, where $E_z = H_z / H_0 = \left[\Omega_{\rm
m} (1+z)^3 + 1 - \Omega_{\rm m} \right]^{1/2}$ (for a flat cosmology
with matter density $\Omega_{\rm m}$ and Hubble constant at the 
present time $H_0$).
Similarly, the definition of the bremsstrahlung emissivity $\epsilon \propto
\Lambda(T) n_{\rm gas}^2 \propto T^{1/2} n_{\rm gas}^2$
(the latter being valid for systems sufficiently hot, e.g. $> 2$ keV)
allows to relate the bolometric luminosity,
$L$, and the gas temperature, $T$:
$L \approx \epsilon R^3 \approx T^{1/2} n_{\rm gas}^2 R^3 
\approx T^{1/2} f_{\rm gas}^2 M_{\rm tot}^2 R^{-3} 
\approx f_{\rm gas}^2 T^2$, where we have made use
of the formula on the $M_{\rm tot}-T$ relation shown
above.  

By combining these basic equations, we can obtain the scaling relations
among the X-ray properties that we are going to investigate in the present
section:
\begin{itemize}
\item $F_z \; M_{\rm tot} \; \propto \; T^{3/2}$
\item $F_z \; M_{\rm gas} \; \propto \; T^{3/2}$
\item $f_{\rm gas} \; \propto \; T^0 \; \propto \; M_{\rm tot}^0$
\item $F_z^{-1} \; L \; \propto \; T^2$
\item $F_z^{-1} \; L \; \propto \; (F_z \; M_{\rm tot})^{4/3} \propto \; (F_z \; M_{\rm gas})^{4/3}$
\item $F_z^{4/3} \; S \propto \; T$,
\end{itemize}
where we have combined all the cosmological dependence in the factor 
$F_z = E_z \times \left( \Delta_z / \Delta_{z=0} \right)^{1/2}$.
All the quantities are measured within regions with overdensity 
$\Delta_z=500$ (see Sect.~2), apart from the entropy $S$ that 
is related to the thermodynamical entropy $K$ according to $K \propto 
\log S$, and is estimated as $T_{\rm gas}(R_{0.1}) / 
n_{\rm e}^{2/3}(R_{0.1})$ with $R_{0.1} = 0.1 \times R_{200}$.

The relative error (at $1 \sigma$ level of confidence) on the physical
quantities is assumed to be 10 per cent for the luminosity, 15 per
cent for the temperature and gas mass and 25 per
cent for the gravitational mass. These uncertainties are the
average values cited presently for local and high$-z$ clusters
(e.g. Finoguenov, Reiprich \& B\"ohringer 2001, Ettori et al. 2004).

The behavior of these scaling laws are examined first 
in their normalization and slope by fitting the logarithmic relation
\begin{equation}
\log Y = \alpha +A \log X
\label{eq:fit}
\end{equation}
between two sets of measured quantities $\{X_j\}$ and $\{Y_j\}$.  We
use the bisector modification (i.e. the best-fit results bisect those
obtained from minimization in vertical and horizontal directions) of
the linear regression algorithm in Akritas \& Bershady (1996 and
references therein, hereafter BCES for Bivariate Correlated Errors and
intrinsic Scatter) that takes into account both any intrinsic scatter
and errors on the two variables considered as symmetric.  We adopt the
best-fit results from the BCES$(Y|X)$ estimator only when the expected
slope is zero, being the inverse one equals to infinity, like in the
cases of the $f_{\rm gas} - T$ relation.
The uncertainties on the best-fit results are obtained from
10\,000 bootstrap resampling.  

With a relevant difference with respect to the analysis presented in 
Paper I, we have selected for the present analysis only  
galaxy clusters with 
$T_{\rm ew}(R_{500}) > 2$ keV. Furthermore, because of the inclusion
of a relative error on the measured quantities, the best-fit
results do not necessarily coincide with those   
provided in Paper I. We refer to the latter work for a detailed
discussion of the (mis)matches between observed and simulated scaling 
relations at $z=0$. We concentrate here on the properties of these
relations at high redshift ($z \ge 0.5$).

In the following sections, we discuss the slope, normalization and
evolution with redshift of the power-law fit among these quantities and 
compare these results to the observational constraints obtained 
by Ettori et al. (2004) for a sub-sample of 22 X-ray galaxy clusters
at $z \ge 0.5$ observed on kpc scales through deep \chandra observations. 
It is worth mentioning that, in the present analysis, we adopt 
systematically the definitions and methods discussed there to study 
the behaviour of the scaling laws of X-ray galaxy clusters. 
The results on the best-fit normalization and slope for the local scaling 
laws here investigated are given in Table~\ref{tab:fit} and shown as 
dotted (when the slope is fixed to the value predicted by the above 
scaling relations) and dashed (when the slope is a free parameter) lines 
in Figures~\ref{fig:lt} -- \ref{fig:st}. 
In Table~\ref{tab:fit2}, a synoptic summary of the best-fit results 
obtained from samples of simulated and observed clusters at $z \ge 0.5$ is 
given, with an assessment on the deviation between the evolution parameter 
$B$ (see Section~4.2) as measured in the simulations and from 
observational data.

\subsection{On the slope and normalization of the scaling relations}

We fit equation~\ref{eq:fit} to the quantities of the simulated 
clusters at $z=0$ and quote the results in Table~\ref{tab:fit}. 
Overall, we observe a slope steeper than predicted
from self-similar model in the examined scaling laws, with $L-T$
(Fig.~\ref{fig:lt}), $M_{\rm tot}-T$ (Fig.~\ref{fig:mt}) and
$M_{\rm gas}-T$ (Fig.~\ref{fig:mgt}) relations
being the ones with deviations larger than $3 \sigma$ (best-fit of $A=
3.3\pm0.3$, $2.1\pm0.2$ and $2.4\pm0.2$, respectively), 
whereas $S-T$ (Fig.~\ref{fig:st}) deviates by about $2.5 \sigma$
from the expected values of $A=1$, $L-M_{\rm tot}$ by $2.2 \sigma$
and $L-M_{\rm gas}$ is consistent with the predicted slope of $4/3$.

%%  forprint,par_ltz[5,*],par_mtz[5,*],par_mgtz[5,*],par_lmz[5,*]
About the normalization of these scaling relations at high redshift 
($z \ge 0.5$) and how they compare to the observed estimates, we use 
the results by Ettori et al. (2004) once the slope is fixed to the 
expected value from simple gravitational collapse assumption.  
The simulated clusters present normalizations that are definitely lower
than what is actually observed in the same redshift range (see 
Table~\ref{tab:fit2}): 
$\alpha$ is lower by 7 per cent in the $M_{\rm tot}-T$ relation, 
12 per cent in the $M_{\rm gas}-T$ relation,
27 per cent in the $L-T$ relation, 
45 per cent in the $L-M_{\rm tot}$ relation, 
18 per cent in the $f_{\rm gas}-T$ relation 
and consistent within few per cent in the $S-T$ relation.  
One possible explanation for these lower normalizations is that
the amount of the hot X-ray emitting plasma measured within $R_{500}$ of
simulated systems at $z \ge 0.5$ is smaller than the observed one.
This deficiency in the density of cosmic baryons detectable as hot gas
in hydrodynamical simulations of galaxy clusters might be complementary 
to the well-known ``over-cooling" problem, where radiative cooling converts
a too large fraction of cosmic baryons into collisionless stars
(e.g. Suginohara \& Ostriker 1998, Balogh et al. 2001, Borgani et al. 2002)
if the local cooling time is not conveniently increased by some sort
of feedback process.
On the other hand, the good agreement in the $L-M_{\rm gas}$ relation
suggests that overestimates of the gas temperature and total mass with
respect to what actually measured can affect our overall results.
We have indeed verified that by reducing, for example, by $\sim$15 per cent the 
simulated estimates of both $T$ and $M_{\rm tot}$
% , as expected when using a 
% spectroscopic-like definition of temperature (e.g. Mazzotta et al. 2004), 
provides a general good match between
the normalizations of the scaling relations in the observed and simulated
datasets. We refer to a future work where we will compare different
definitions of gas temperatures, as inferred from simulations,
to observational data, and the implications on scaling
relations and measurements of the cluster mass.

\begin{figure}
 \epsfig{figure=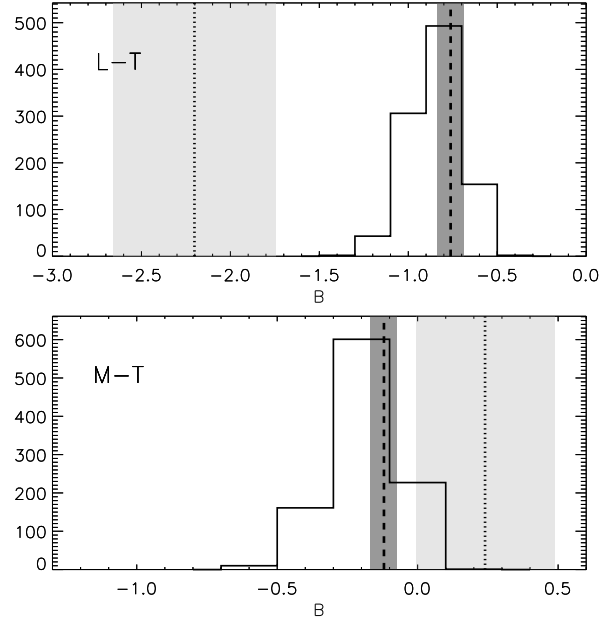,width=0.5\textwidth}
\caption{Distribution of the values of the evolution parameter $B$ as
measured in sub-samples of 22 simulated objects to mimic the dataset
at $z \ge 0.5$
in Ettori et al. (2004). {\it Dark shaded region:} $1-\sigma$ range
and best-fit ({\it dashed line}) of $B$ as evaluated for the overall
sample of simulated clusters. {\it Light shaded region:} $1-\sigma$ range
and best-fit ({\it dotted line}) of $B$ as evaluated in Ettori et al. (2004)
for the same relation.
} \label{fig:b_mc}
\end{figure}

\subsection{On the evolution of the scaling laws}

We measure a median gas temperature of 2.5, 2.4, 2.3 and 2.3 keV in our
samples at redshift of $0$, $0.5$, $0.7$ and $1$, respectively. 
On the other hand, we note a drastic decrease in the median values of 
$F_z^{-1} L / (10^{44}$ erg s$^{-1})$ from 1.01 at $z=0$ to 0.61, 0.55 and 0.42 at 
$z=0.5, 0.7$ and $1$, mostly due to the cosmological correction introduced
by the evolution of the Hubble constant and parametrized through $E_z$.
Significantly lower values than those measured locally are also evaluated
for the quantity $F_z M_{\rm tot} / (10^{14} M_{\odot})$ 
that changes from 1.94 ($z=0$) to 1.86 ($z=0.5$), 1.63 ($z=0.7$) and 1.53 ($z=1$).
These simple calculations give already indications for a declining
luminosity and mass for given temperature and with increasing redshift or,
in other words, for a {\it negative} evolution of the corresponding 
scaling relations.
 
To constrain properly the evolution in the normalization of the
considered scaling laws, we proceed as discussed in Ettori et
al. (2004). First, we fix $(\alpha, A)$ to the best-fit results
obtained from the clusters at $z=0$, $(\overline{\alpha},
\overline{A})$, and, then, evaluate the confidence interval through a
least-square minimization on the parameter $B$ in the relation
\begin{equation}
\log Y = \overline{\alpha} +\overline{A} \log X +B \log (1+z).
\label{eq:bfit}
\end{equation}
In particular, for a given grid of values of $\{B_i\}$, we search for the
minimum of the merit function
\begin{equation}
\chi^2_i = \sum_j \frac{\left[ \log Y_j -\overline{\alpha} -\overline{A} 
\log X_j -B_i \log (1+z_j) \right]^2}{\epsilon_{\log Y_j}^2 
+\epsilon_{\overline{\alpha}}^2 +\overline{A}^2 \epsilon_{\log X_j}^2
+\epsilon_{\overline{A}}^2 \log^2 X_j}, 
\label{eq:chi2}
\end{equation}
where the errors on the best-fit local values,
$\epsilon_{\overline{\alpha}}$ and $\epsilon_{\overline{A}}$,
are considered and propagated with the uncertainties,
$\epsilon_{\log X} = \epsilon_X/(X \, \ln10)$ and $\epsilon_{\log Y} =
\epsilon_Y/(Y \, \ln10)$, on the measured quantities.

We measure a small, but highly significant, {\it negative} (i.e. $B<0$) 
evolution in the $L-T$ ($B \approx -0.8$) and $L-M_{\rm tot}$ 
($B \approx -0.6$) relations. Marginally ($<3 \sigma$) negative values of $B$
are measured in the $M_{\rm tot}-T$, $M_{\rm gas}-T$ and $f_{\rm gas}-T$ 
relations. A remarkable positive evolution is estimated in the
$S-T$ relation. 
All these trends becomes more significant when objects at higher
redshift are considered. 
In particular,  
the relative differences between local and $z=1$ objects have been plotted 
in the lower panels of Figures~\ref{fig:lt} -- \ref{fig:st}, where the 
corresponding physical quantities are referred to the best-fit local 
power-laws. A clear segregation between local and high$-z$ clusters is 
present in, e.g., the $L-T$ (Fig.~\ref{fig:lt}) and $L-M_{\rm tot}$ 
(Fig.~\ref{fig:lm}) plots.

%%%%%%%%%%%%%%%%%   check  vs.  OBSERVATION
All these values in the evolution of the studied scaling relations 
are qualitatively consistent with the recent observational
constraints (e.g. Ettori et al. 2004 and see Table~\ref{tab:fit2}).
On the other side, the amount of measured evolution is less than actually 
observed in the simulated $L-T$, $L-M_{\rm tot}$ and $S-T$ relations. 

To check against any bias effect due to the number of objects selected
in the sample, we have randomly extracted 7 objects at $z=0.5$, 10 at
$z=0.7$ and 5 at $z=1$ to mimic the distribution of the 22 observed
clusters at $z \ge 0.5$ in Ettori et al. (2004) in which 7 clusters have
redshift between $0.5$ and $0.65$, 10 are between $z=0.65$ and $z=0.85$
and 5 have $z$ between $0.85$ and $1.3$.  We have repeated the sampling
for 1,000 times and plot the distribution of the values of the
evolution parameter $B$ in fig.~\ref{fig:b_mc}.  We conclude that
no-significant deviations from the best-fit $B$ value as measured for
the overall sample are obtained when sub-samples are considered.
Moreover, the observed evolution in the $L-T$ relation cannot be
reproduced at very high significance ($>99.9$ per cent), whereas the
one observed in the $M_{\rm tot}-T$ relation is only marginally
consistent with the distribution of the $B$ values (about 23 per cent
of the simulated sub-samples have an evolution parameter that falls
within the $1-\sigma$ error range of the observed $B$ value).

\subsection{On the results obtained by adopting different selection criteria}

We have repeated the analysis discussed above by extracting the simulated
objects according to different criteria to answer two main issues:

\begin{enumerate}
\item {\it What happens if we select only very massive systems
(e.g. $T_{\rm ew} >3$ keV) in accordance to the observational datasets
available ?}  By construction, no very massive virialized systems are
build in this cosmological simulations implying that a selection at
higher threshold in temperature (mass) reduces significantly the
number of objects available. When we select the simulated clusters
with $T_{\rm ew} >3$ keV, we collect just 24 objects at $z=0$ and 8 at
$z=1$ (14 and 12 at $z=0.5$ and $0.7$, respectively).  Therefore, by
decreasing the number of objects, we increase the statistical
uncertainties of the best-fit parameters and no more significant
evolution are detected. For example, in the $L-T$ relation we
constrain the local slope to be $3.4 \pm 1.6$, with a marginally
negative evolution at $z \ge 0.5$ ($B \approx -0.7 \pm 0.4$).  An
evolution consistent with zero is measured in the $M_{\rm tot}-T$
relation ($B = -0.1 \pm 0.3$ at $z \ge 0.5$), which presents a slope
at $z=0$ consistent with the predicted value of 1.5 ($2.1 \pm 1.1$).

\item {\it How do the relations change when the gas mass-weighted
temperature, $T_{\rm mw}$, is used instead of the emission-weighted
one (see, e.g., Allen et al. 2001, Thomas et al. 2002) ?}  If we
select simulated clusters with $T_{\rm mw} > 2$ keV, we reduce
significantly the number of objects selected (42 vs. 97 simulated
clusters at $z=0$ and 12 vs. 72 at $z=1$) and confirm by an indirect
way the well-recognized problem of simulated cores being hotter than
actually observed (see discussion on the temperature profile in
section~3.4 in Paper I).  All the scaling relations appear to be
flatter and more consistent with self-similar predictions than 
measured when $T_{\rm ew}$ is in use: e.g., slopes of $3.0 \pm 0.6$
and $1.6 \pm 0.2$ are measured in the local $L-T$ and $M_{\rm tot}-T$
relations.  Also the evolution parameter $B$ departs less
significantly from zero, with values that are typically 50 per cent
lower than the measured ones in the original selection.
 
% \item {\it How much improvement in the detection of the evolution is obtained
% as a consequence of a decrease in the relative observational errors
% (let assume here by 50 per cent) ?} 
% We have assumed relative errors (at $1 \sigma$ level of confidence) of
% 10 per cent on the luminosity, 15 per cent on the measurements of the 
% temperature and gas mass and 25 per cent on the gravitational mass estimates.
% These amounts are the average values measured nowadays in local and
% high$-z$ clusters (e.g. Finoguenov et al. 2001, Ettori et al. 2004). 
% By reducing these amounts by half, all the statistical constraints become 
% tighter with an increased significance in assessing any evolution.  
% For example, $B$ is measured as $11 (13) \sigma$ deviation from zero in the 
% $L-T$ ($M_{\rm tot}-T$) relation, instead of a departure significant at 
% $4 (7) \sigma$ with the assumed errors at $z=1$. 
\end{enumerate}

\section{Summary and discussion}

Using a large cosmological hydrodynamical simulation, we have selected
galaxy clusters at redshift $z=0,~0.5,~0.7$ 1, all having $T_{\rm ew}$
within $R_{500}$ larger than $2$ keV, to explore the distribution of
baryons and the X-ray scaling relations as function of redshift.  We
adopt an approach in our analysis that mimics observations,
associating with each measurement an error comparable to recent
observations and providing best-fit results obtained with robust
techniques.

Our main findings can be summarized as follows:

\begin{enumerate}
\item At $z \ge 0.5$, the simulated gas temperature and density profiles
tend to show more scatter around the mean value, suggesting that many
systems are still in formation and dynamically younger than the ones
simulated locally.  The temperature and density gradients are flatter
and less centrally peaked, respectively, than those at $z=0$.
However, the overall shape of these radial profiles represented by the
polytropic index does not change significantly.

\item Gas density profiles for simulated clusters are slightly steeper
  that actually
observed at $z \approx 1$, with deviations of about 10-20 per cent (on average)
below $0.1 \times R_{500}$ and above $0.7 \times R_{500}$.
This implies a mean integrated value of the simulated gas mass lower by about 
10 per cent.
A significantly larger scatter in the observed data suggests that
either the complex dynamical history of the formation of these structures
is not recovered fully in this hydrodynamical simulation or the
simulated dark matter halos are more regular due to the criteria adopted in 
their selection.

\item The baryons within the virial radius are distributed among a
cold phase, with a relative contribution that increases from less than
1 to 3 per cent at higher redshift, a collisionless phase in stars of
about 20 per cent and the X-ray emitting plasma that contributes by 80
(76) per cent at $z=0 (1)$ to the total baryonic budget.  A depletion of
cosmic baryons of the order of $\sim$7 per cent is measured locally at
the virial radius, $R_{\rm vir}$. The average value of the depletion
decreases slightly to 5 per cent at $z=1$.  These values are
consistent with the results obtained in adiabatic hydrodynamical
simulations, even though they have to be considered as lower limits
considering the measured over-production of stars.

\item The X-ray emitting gas mass fraction, $f_{\rm gas}(<r)$,
increases with $r$ as a power-law with a characteristic scale that
also increases from 0.28 to 0.34 $\times R_{\rm vir}$ when going from
$z=0$ to $z=1$, with a typical slope of about 0.25. The measured gas
fraction in the ten hottest systems at $R_{2500}$, $R_{500}$ and
$R_{\rm vir}$ is 0.62, 0.73 and 0.76 $\times (\Omega_{\rm
b}/\Omega_{\rm m})$ at $z=0$ and 0.57, 0.70 and 0.73 $\times
(\Omega_{\rm b}/\Omega_{\rm m})$ at $z=1$, respectively.  The
differences as function of redshift are not significant being within
the measured scatter.

\item We confirm that, also at high$-z$, simulated clusters have
X--ray scaling relations between temperature, $T$, luminosity, $L$,
central entropy, $S$, gas mass, $M_{\rm gas}$, and total gravitating
mass, $M_{\rm tot}$, which are steeper than predicted by simple
gravitational heating.

\item When we fix the slope of these scaling laws to the values
expected from the self-similar scenario, normalizations $\alpha$ lower
by 10--40 per cent are measured in the $L-T$, $M_{\rm tot}-T$, $M_{\rm
gas}-T$ and $L-M_{\rm tot}$ relations at $z \ge 0.5$ when compared to
the observational results for high redshift clusters (e.g. Ettori et
al. 2004). The normalization of the central entropy -- temperature
relation is instead higher by few per cent, confirming the lack of gas
mass in the central regions of these high$-z$ simulated systems.  
On the other hand, the good agreement in the $L-M_{\rm gas}$ relation
suggests that overestimates of the gas temperature and total mass with
respect to what actually measured can affect our overall results.
We have indeed verified that by reducing by 15 per cent the simulated estimates
of both $T$ and $M_{\rm tot}$ provides a general good match between
the normalizations of the scaling relations in the observed and simulated
clusters.
In this perspective, it should be noticed that Mazzotta et al. (2004) have
recently suggested a spectroscopic-like definition of the ICM
temperature from simulations that better matches the temperature
obtained from the spectral fit of observational data.  
This alternative definition tends to provide temperature estimates lower
than the emission-weighted values when the thermal structure of the cluster
in exam deviates more from an isothermal modelization.

\item We measure a declining luminosity and mass for given temperature
and with increasing redshift. This {\it negative} evolution is
parametrized through a power-law dependence upon the redshift,
$(1+z)^B$, with $B \approx -0.8$ for the $L-T$ relation, $B \approx
-0.2$ for the $M_{\rm tot}-T$ relation, $B \approx -0.2$ for the
$M_{\rm gas}-T$ relation and $B \approx -0.6$ for the $L-M_{\rm tot}$
and $L-M_{\rm gas}$ relations.  A slightly positive evolution of $B
\approx +0.2$ is measured in the $S-T$ relation.  As shown in
Table~\ref{tab:fit2}, these results are consistent (deviation less
than $3 \sigma$) with the observed evolution evaluated in a similar
redshift range (e.g. Ettori et al. 2004) for the $M_{\rm tot}-T$,
$M_{\rm gas}-T$ and $f_{\rm gas}-T$ relations. Significantly less
evident evolution are instead measured in the simulated $S-T$,
$L-M_{\rm tot}$ and $L-T$ relations.  We have verified that the amount
of evolution estimated here is robust against bias effects due to the
number of objects examined.
% \item These results on the scaling relations allow to make some predictions
% on the status of very high$-z$ galaxy clusters.
% E.g., a hot ($T_{\rm ew}>2$ keV) system is expected to have an observed
% bolometric luminosity of $\sim 10^{42.8} T_{\rm ew}^{2.9} E_z (1+z)^{-0.6} 
% h_{70}^{-2}$  erg s$^{-1}$ $\propto (1+z)^{0.5}$ in the regime $z \ga 2$
% with a typical flux of few times $10^{-16}$ erg s$^{-1}$ cm$^{-2}$.
\end{enumerate}

Our results demonstrate that the present dataset of simulated
galaxy clusters tends to evolve at higher redshift, with a flatter 
gas temperature profile and a less centrally peaked density profile
and a statistically significant deficiency of X-ray emitting plasma
(items i, iii, iv and vii above).

The comparison with the observational constraints at $z \ge 0.5$, 
performed by adopting the same definitions of the examined quantities
and analyzing comparable samples both in size and physical properties, 
further indicates that the measured lower normalizations (item vi above)
suggest either a significant lack of X-ray luminous baryons
or an overestimates of the simulated values of the gas temperatures and
total gravitating mass. 
An evolution toward less luminous and massive systems for given gas
temperature at higher redshift (item vii) is measured in the 
simulated dataset in good agreement with recent observational 
results. 

Overall, our results show how the study of the evolution in the scaling
relations can give indications on the properties of the
feedback that can be investigated through extensive hydrodynamical
simulations.  
In particular, these results support the request for
either more efficient feedback or extra physical processes, which are
able to suppress (i) the over production of stars in the central
cluster regions and (ii) the lack of cosmic baryons in the hot X-ray
emitting phase.  Moreover, this additional source of heating is
required not only to provide extra energy but also to distribute it
radially in such a way (iii) to improve the agreement between the
simulated and observed gas temperature both locally and up to redshift
of $1$, (iv) to reduce the steepening of the gas density profile at
high$-z$, (v) to `soften' the gas density profiles of low temperature
objects by increasing the central entropy, as already discussed in
Paper I. 
%SB.
%In this perspective, the modelization and simulation of the
%central cluster regions at higher density deserve efforts and
%particular care since the interplay between radiative cooling and
%feedback heating is of crucial relevance to explain the thermal
%structure and the observed properties of the X-ray emitting
%intracluster medium.

The work presented here demonstrates that we are approaching an age in
which code efficiency and supercomputing capabilities make the
simulations able to describe the formation and evolution of cosmic
structures over a fairly large dynamical range in a fruitful symbiosis
with the present observational limits.  The real challenge for
improving the exchange between numerical and observational cosmology
is, on the one side, to build algorithms that more faithfully
incorporate all those astrophysical processes that are suspected to
give relevant contribution to the observational properties of highly
non-linear and over-dense structures, and on the other side, to
resolve spatially with a sufficient number of photons even at $z
\approx 1$ all the physical quantities of interest, starting with the
temperature of the X-ray emitting plasma.

\section*{ACKNOWLEDGEMENTS} 
We thank the anonymous referee for helpful comments that improved the
presentation of the work.
The results presented in this paper are
based on a simulation realized using the IBM-SP4 machine at the
``Consorzio Interuniversitario del Nord-Est per il Calcolo
Elettronico'' (CINECA, Bologna), with CPU time assigned thanks to the
INAF--CINECA numerical Key-Project ``{\it A Tree+SPH High-Resolution
Simulation of the Cosmic Web}'' (2003).
This work has been partially supported by the INFN--PD51 grant and by MIUR.

\end{document}